\documentclass[fleqn,usenatbib]{mnras}
\usepackage{newtxtext,newtxmath}

\usepackage[T1]{fontenc}

\DeclareRobustCommand{\VAN}[3]{#2}
\let\VANthebibliography\thebibliography
\def\thebibliography{\DeclareRobustCommand{\VAN}[3]{##3}\VANthebibliography}

\usepackage{graphicx}	
\usepackage{amsmath}	

\usepackage{epstopdf}

\title[Chemical abundance variations in young star clusters]{Searching for chemical abundance variations in young star clusters in the Magellanic Clouds: NGC 411, NGC 1718 and NGC 2213}

\author[Shalmalee Kapse, Richard de Grijs and Daniel B. Zucker]{
Shalmalee Kapse$^{1,2}$\thanks{E-mail: shalmalee.kapse@students.mq.edu.au},
Richard de Grijs$^{1,2,3}$ and
Daniel B. Zucker$^{1,2}$
\\
$^{1}$Department of Physics and Astronomy, Macquarie University, Balaclava Road, Sydney, NSW 2109, Australia \\
$^{2}$Research Centre for Astronomy, Astrophysics and Astrophotonics, Macquarie University, Balaclava Road, Sydney, NSW 2109, Australia \\
$^{3}$International Space Science Institute–-Beijing, 1 Nanertiao, Zhongguancun, Hai Dian District, Beijing 100190, China
}

\date{Accepted XXX. Received YYY; in original form ZZZ}

\pubyear{2020}

\begin{document}
\label{firstpage}
\pagerange{\pageref{firstpage}--\pageref{lastpage}}
\maketitle

\begin{abstract}
The conventional picture of coeval, chemically homogeneous, populous star clusters -- known as `simple stellar populations' (SSPs) -- is a view of the past. Photometric and spectroscopic studies reveal that almost all ancient globular clusters in the Milky Way and our neighbouring galaxies exhibit star-to-star light-element abundance variations, typically known as `multiple populations' (MPs). Here, we analyse photometric {\sl Hubble Space Telescope} observations of three young ($<$2 Gyr-old) Large and Small Magellanic Cloud clusters, NGC 411, NGC 1718 and NGC 2213. We measure the widths of their red-giant branches (RGBs). For NGC 411, we also use a pseudo-colour--magnitude diagram (pseudo-CMD) to assess its RGB for evidence of MPs. We compare the morphologies of the clusters' RGBs with artificially generated SSPs. We conclude that their RGBs do not show evidence of significant broadening beyond intrinsic photometric scatter, suggesting an absence of significant chemical abundance variations in our sample clusters. Specifically, for NGC 411, NGC 1718 and NGC 2213 we derive maximum helium-abundance variations of $\delta Y=0.003\pm0.001 (Y=0.300), 0.002\pm0.001 (Y=0.350)$ and $0.004\pm0.002 (Y=0.300)$, respectively. We determined an upper limit to the NGC 411 nitrogen-abundance variation of $\Delta$[N/Fe] = 0.3 dex; the available data for our other clusters do not allow us to determine useful upper limits. It thus appears that the transition from SSPs to MPs occurs at an age of $\sim$2 Gyr, implying that age might play an important role in this transition. This raises the question as to whether this is indeed a fundamental minimum-age limit for the formation of MPs.
\end{abstract}




\section{Introduction}

Star clusters were traditionally thought to represent `simple' stellar populations (SSPs), that is, all cluster member stars were assumed to have formed at approximately the same time, with the same chemical composition, so that they can be described by a single isochrone in their colour--magnitude diagrams \citep[CMDs;][]{Vesperini2012DynamicsClusters,Cabrera-Ziri2016No411,Bastian2013ConstrainingClusters,Li2016StellarClusters,Bastian2013ConstrainingClusters}. However, analyses of some old globular clusters (GCs; with ages $>$10 Gyr) have progressively challenged the SSP approximation. For example, current observations support the idea that nearly all GCs host multiple populations (MPs), either displaying multiple main sequences \citep[MSs;][]{Piotto_2007}, sub-giant branches \citep[SGBs;][]{Anderson2009MixedTUC}, red-giant branches \citep[RGBs;][]{Piotto2007ObservationalClusters,681408180,675688537} and/or horizontal branches \citep[HBs;][]{675688503}. In addition, an increasing body of combined photometric \citep{Martocchia2018TheNGC1978} and spectroscopic \citep{2018AJ....156..116M} analyses  of massive clusters also supports the presence of internal chemical abundance variations at much younger ages.

Nevertheless, it has been established conclusively that the SSP approximation still holds for at least a few star clusters in nearby galaxies. A number of young ($\sim$1--3 Gyr-old) massive clusters (with minimum masses of a few $\times 10^{4}$ M$_\odot$) do not exhibit measurable internal chemical abundance spreads \citep[e.g.,][]{Mucciarelli2010ChemicalCloud,Cabrera-Ziri2016No411}. Indeed, rapid star formation has been hypothesised to naturally lead to the ejection of some stars and most residual gas from young clusters \citep{675688500,675688501,675688516}, a process for which some observational evidence has been obtained in a number of extremely young and massive clusters \citep{675688498,675688513,Bastian2013ConstrainingClusters}. 

Several scenarios have been proposed which advocate cluster formation through an initial burst of star formation \citep[e.g.,]{2013MNRAS.429.1913V,675688545}. Residual intracluster gas, polluted by the ejecta from the first generation of stars, is thought to provide fuel for the formation of a second stellar generation. The latter could potentially be identified based on the distributions of stellar chemical abundances within a cluster. This scenario is, in fact, one of the most popular hypotheses proposed for star cluster formation. Various polluters could contribute to the formation of a second stellar generation, including young asymptotic giant-branch (AGB) stars \citep{681408158}, rotating massive stars \citep{675688511}, interacting binary systems \citep{675688497} and supermassive stars \citep{2014MNRAS.437L..21D}.

A possible alternative scenario for MP formation has been suggested \citep{675688504,Li2016StellarClusters}, involving either external accretion or cluster mergers \citep{675688545}. Nearly all Milky Way GCs are older than 10 Gyr, and hence they cannot be used to test the possible age dependence of MPs as important driver of their origin. Only a very small number of Milky Way clusters with ages younger than $2 \times 10^8$ yr are sufficiently massive ($> 10^4$ M$_\odot$) to be useful for investigations of the emergence of MPs (e.g., NGC 2121, NGC 2155, NGC 2193, Kron 3). In nearby galaxies such as the Large and Small Magellanic Clouds (LMC, SMC), a well-known cluster age gap is present, with a paucity of observed clusters between approximately 2 and 6 Gyr \citep{675688495}. At an age of less than $\sim$2 Gyr, NGC 1978 is the youngest star cluster known to show the presence of MPs \citep{675688527}. If we consider age as one of the driving factors for MP formation, then the transition of SSPs to MPs would appear to commence at $\sim$2 Gyr \citep{Li2016StellarClusters}. 

Given that most young, massive clusters appear consistent with being SSPs, whereas many old GCs exhibit star-to-star chemical dispersions, a number of interesting questions arise. The fundamental issue remains as to whether these dispersions represent true MPs or whether they are perhaps indicative of variations in the stellar models describing SSPs. Cluster mass has been established as one of the dominant factors determining which clusters host apparent MPs \citep{Li2014Not-so-simple1868}. This is not surprising, because if one adopts the prevailing primary hypothesis for the production of MPs, i.e., collisions of the stellar winds from AGB stars, one would expect MPs to form only in massive star clusters in the first place. Such clusters must be sufficiently massive to retain the AGB ejecta in their gravitational potential wells. Another important, still open question is whether there is a lower-age limit for clusters to host MPs. In other words, what is the age range at which SSPs tend to become MPs?  Therefore, here we address the key question as to whether age is a key parameter in the formation of MPs, alongside cluster mass. 

Recent surveys of massive clusters in the Magellanic Clouds have managed to fill in the 2--6 Gyr cluster age gap. To further explore old stellar populations, measurements of chemical abundance variations have proved decisive \citep[e.g.,][]{675688519}; a number of spectroscopic studies of chemical abundance variations in old GCs have revealed sodium--oxygen (Na--O) and aluminium--magnesium (Al--Mg) anti-correlations, especially among bright stars \citep{675688519}. Many old GCs show multiple sequences in their CMDs, in turn implying the presence of chemical variations \citep{675688526}. Broader widths of such sequences could be another indication of partially hidden abundance variations. Despite these promising insights and developments in our understanding of old GCs, similar analyses of clusters in the young (1--2 Gyr) and intermediate-age (2--6 Gyr) ranges have yet to be undertaken in a systematic fashion.

Two of the most common elements that may give rise to multiple sequences in a cluster's CMD are helium (He) and nitrogen (N) \citep{2018ARA&A..56...83B}. Multiple sequences can also be discovered spectroscopically, thus supporting photometric analyses. Stellar optical luminosities are affected by He abundance variations \citep{675688519}, whereas ultraviolet (UV) luminosities are affected by N abundance variations \citep{681408180,Milone2015ThePopulations}. A combination of observations in optical and UV passbands can thus provide a much better understanding of chemical abundance variations based on the CMDs of star clusters. 

To establish a lower limit to the age of clusters at which MPs start to occur, it is important to investigate young and intermediate-age clusters; such clusters are most readily observable in the LMC and SMC. With this aim in mind, we have selected two LMC clusters (NGC 1718 and NGC 2213) and one SMC cluster (NGC 411) with ages younger than $\sim$2 Gyr (see below). Our goal is to constrain the maximum chemical abundance spread in these clusters based on a detailed analysis of their RGBs. NGC 411, NGC 1718 and NGC 2213 have ages of 1.8, 1.4 and 1.78 Gyr, respectively \citep{10.1093/mnras/stw1491,Goudfrooij_2014}. We investigate these clusters using the same methodology as that developed by \cite{Zhang2018No1783}, in essence comparing their RGB morphologies with those of SSPs composed of artificial stars (see below).  
\begin{figure}
    \centering
    \includegraphics[width=0.5\textwidth]{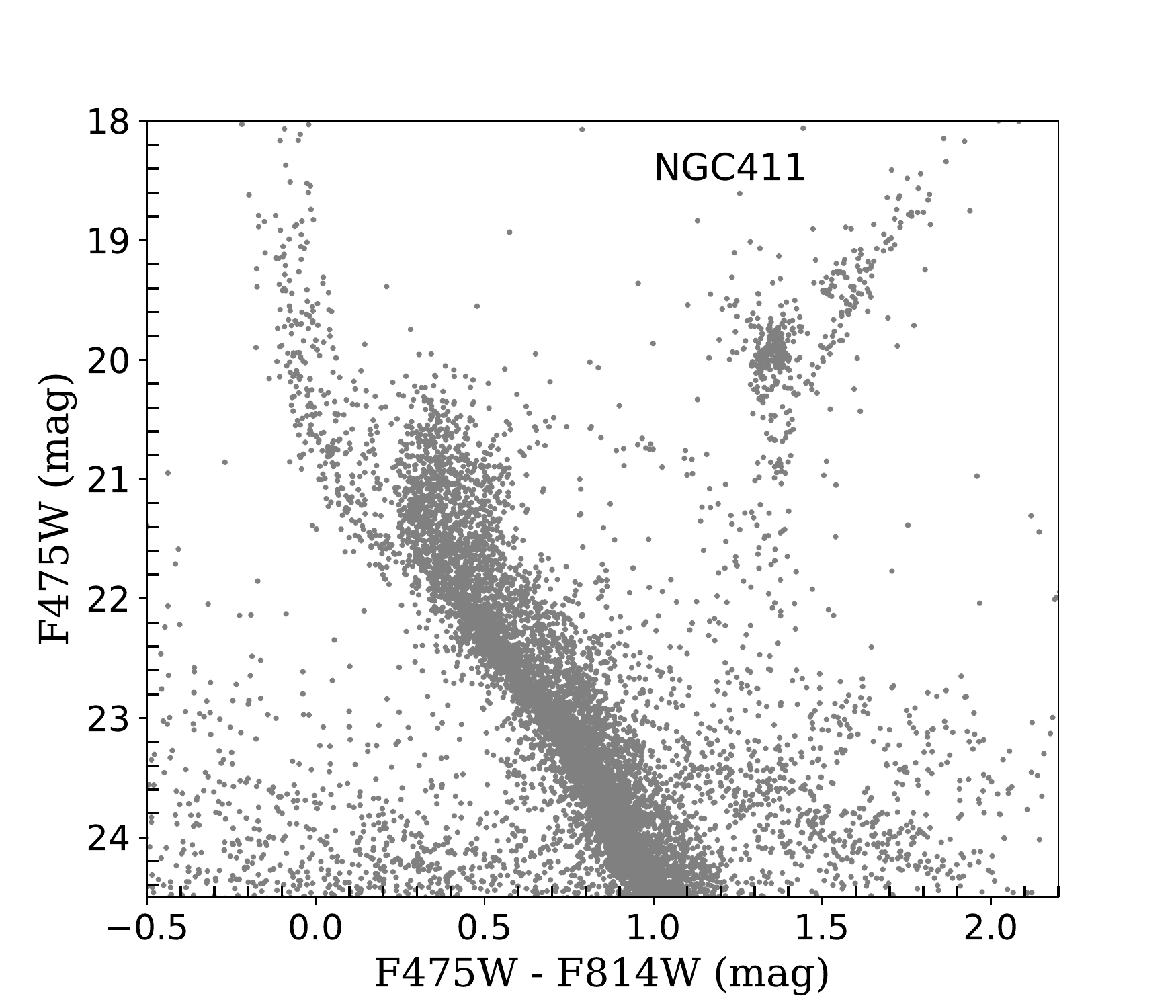}
    \includegraphics[width=0.5\textwidth]{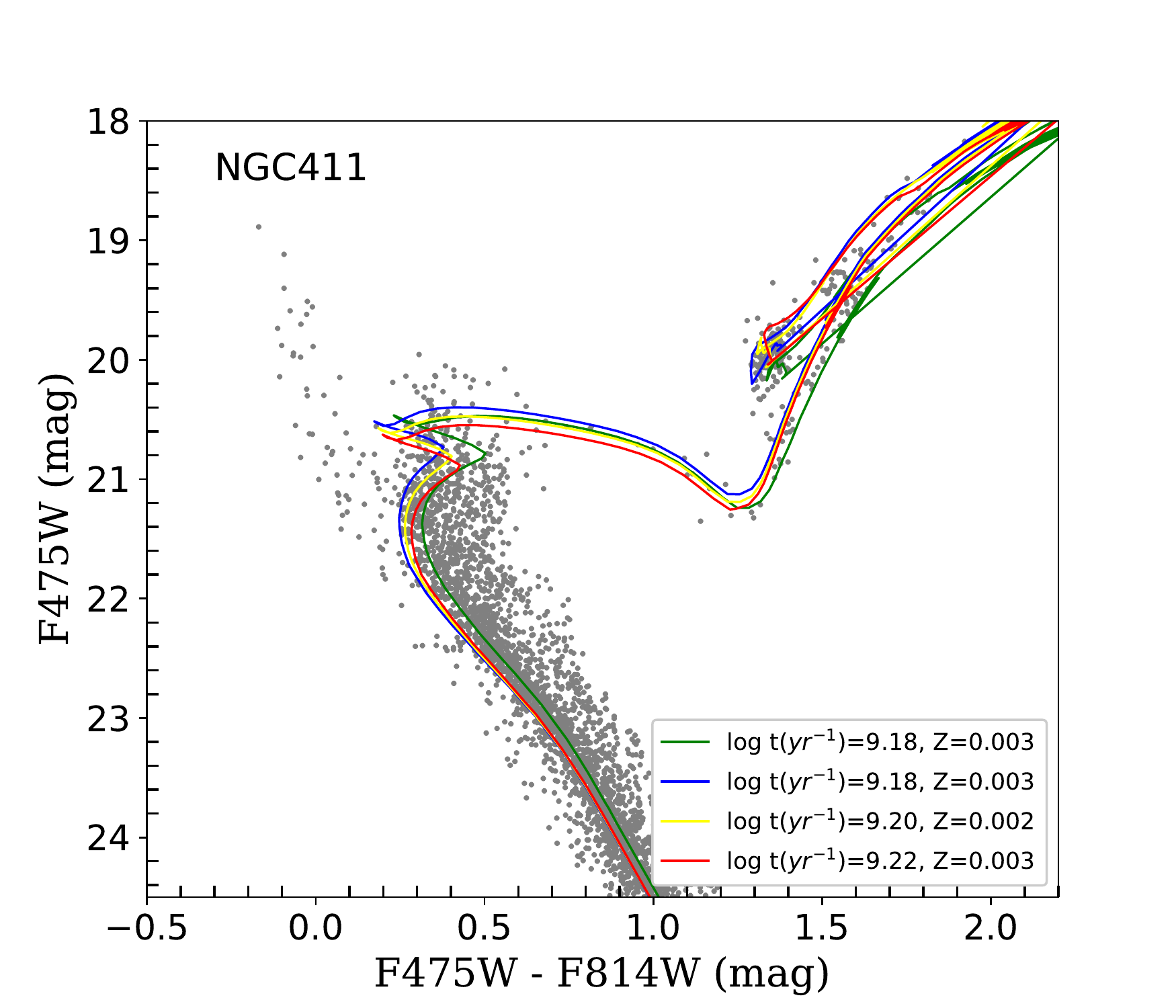}
    \caption{NGC 411 CMDs. (top) Raw data. (bottom) Field-star-decontaminated CMD. The isochrone parameters pertaining to the bottom panel are log({\it t} yr$^{-1}$) = 9.18 ($t \sim 1.5$ Gyr), $(m-M)_0=18.5$ mag, $A_{V} = 0.25$ mag and $Z=0.003$.}
\end{figure}
Thus far, NGC 1978 is the only cluster known to display MPs at an age younger than 2 Gyr. Clusters younger than $\sim$2 Gyr remain poorly investigated. A few such clusters were studied many years ago. For instance, photometric results for NGC 1718 were first presented by \cite{refId1}, some 19 years ago. Although NGC 2213 has been studied more recently by \cite{10.1093/mnras/sty580}, these authors focussed more on the cluster's blue straggler stars than on its RGB stars. Finally, the sub-giant branch of NGC 411 has recently been studied by \cite{10.1093/mnras/stw1491}.

%

\begin{figure*}
\centering
\includegraphics[width=.4\textwidth]{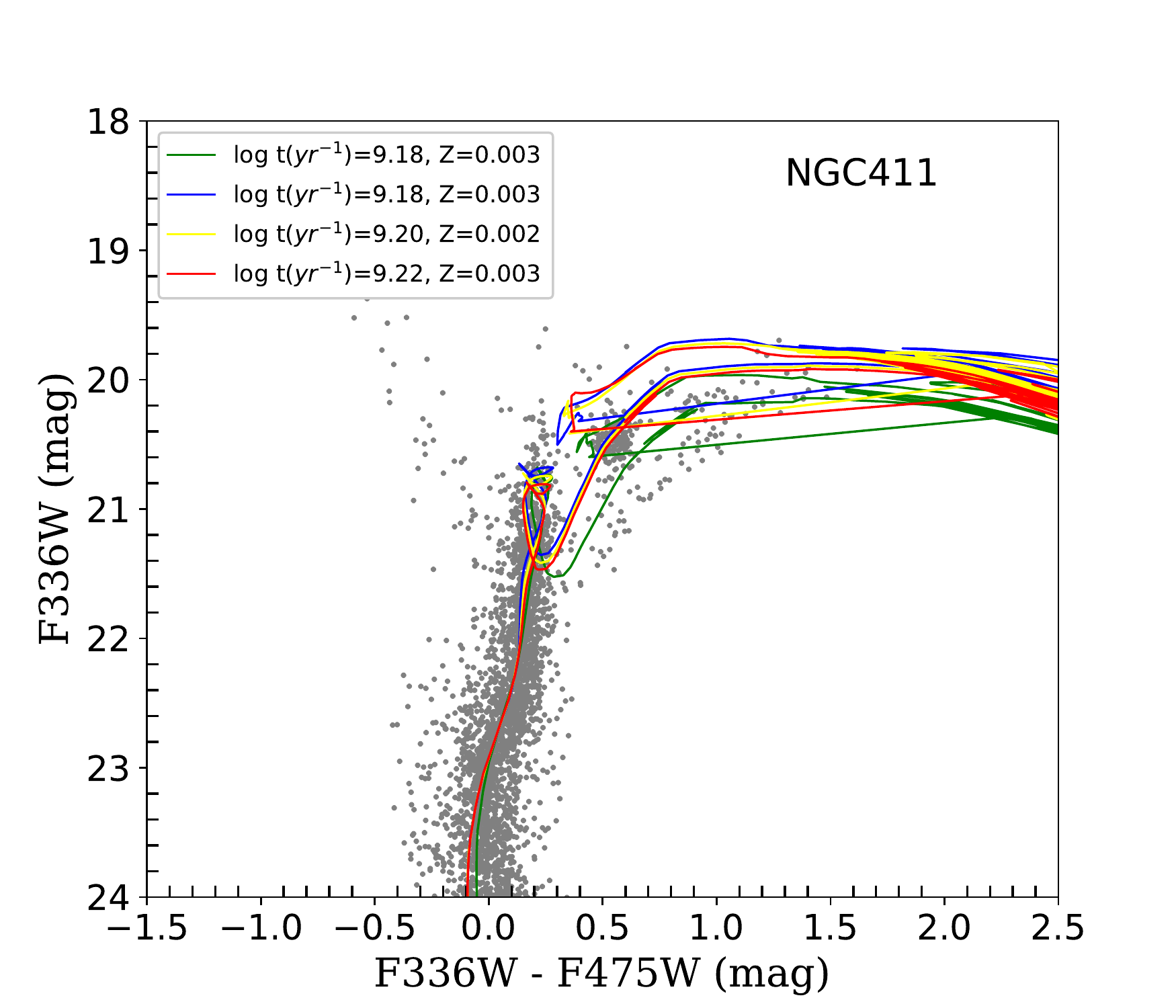}\quad
\includegraphics[width=.4\textwidth]{NGC_411_475_814.pdf}\quad
\includegraphics[width=.4\textwidth]{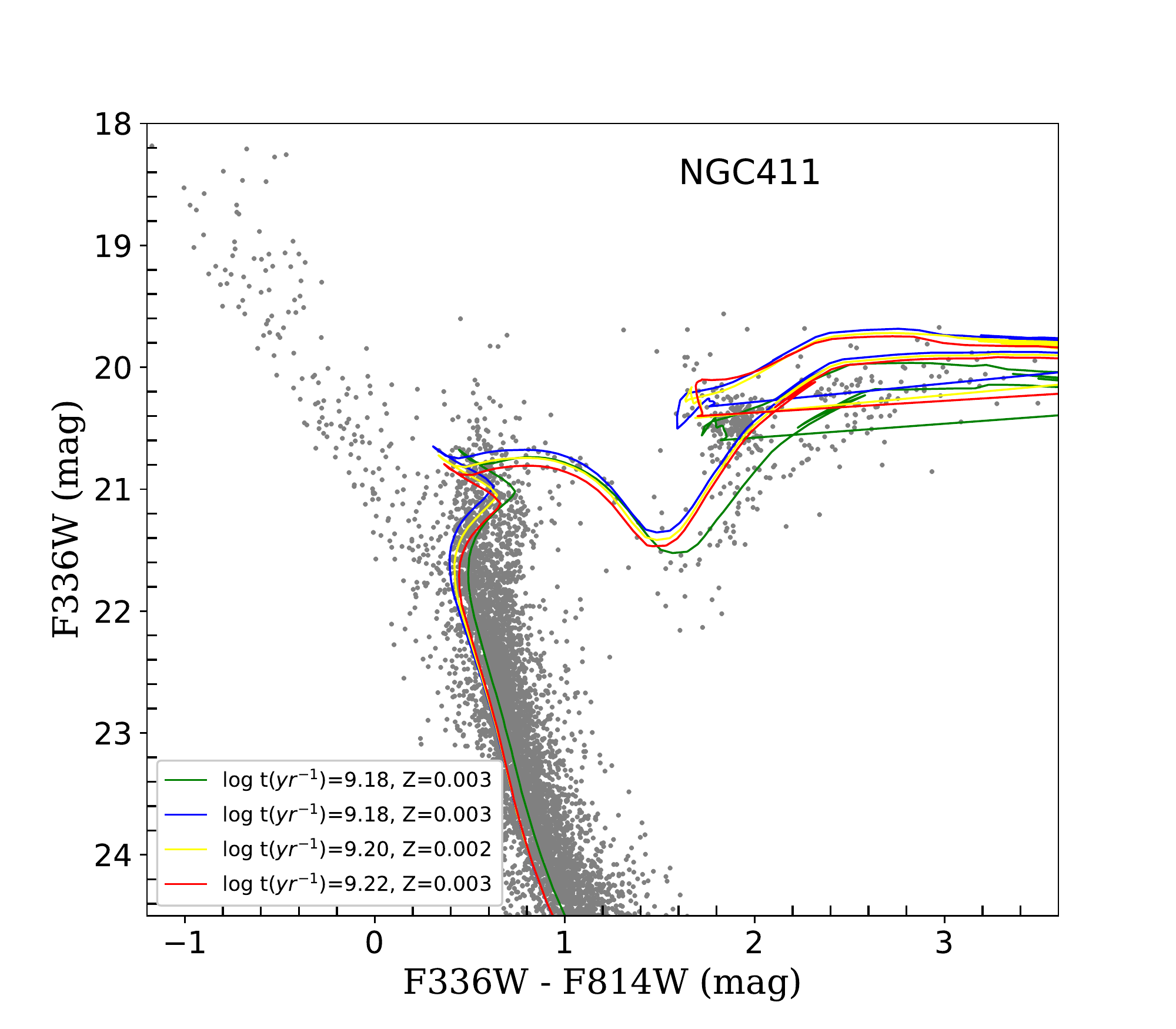}
\medskip
\includegraphics[width=.4\textwidth]{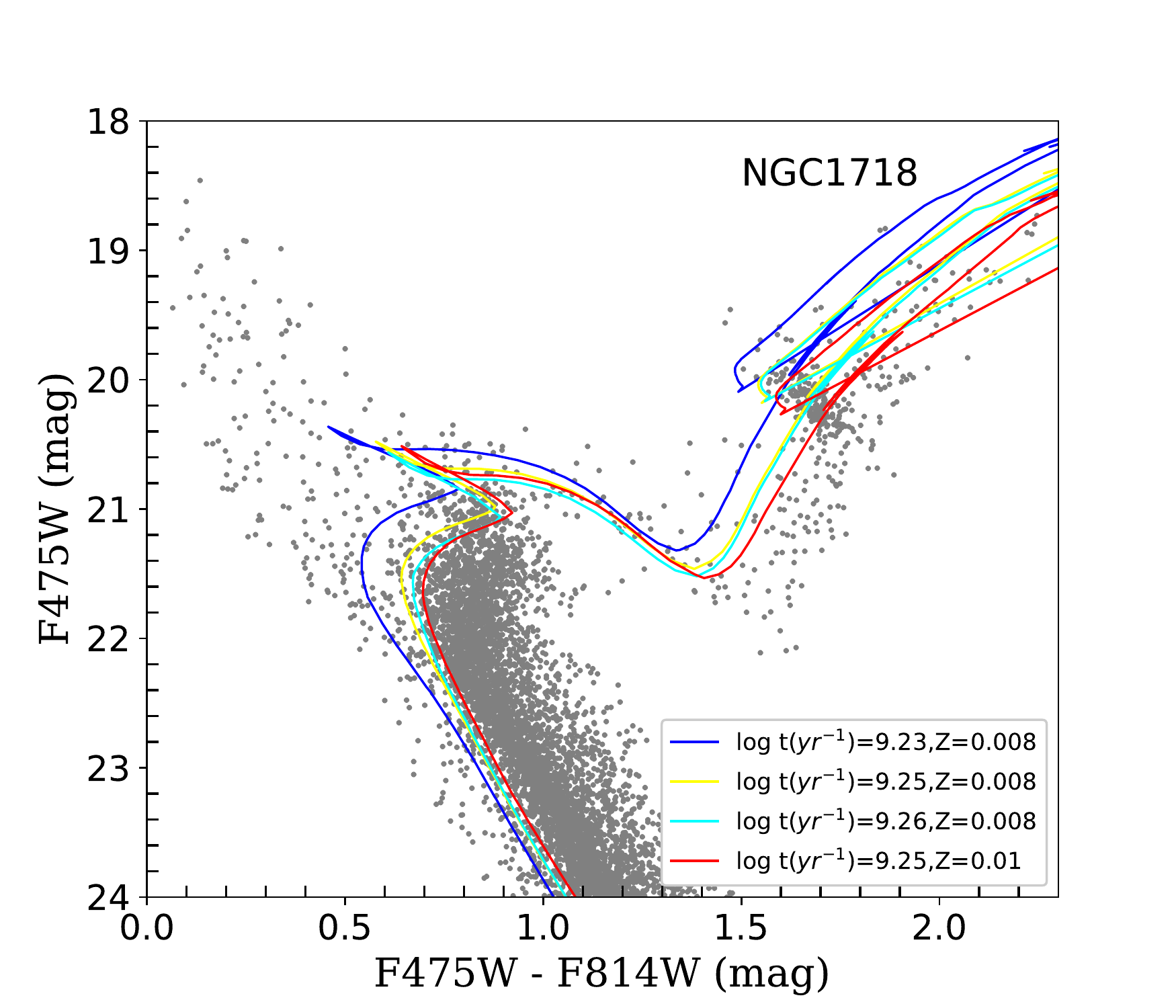}\quad
\includegraphics[width=.4\textwidth]{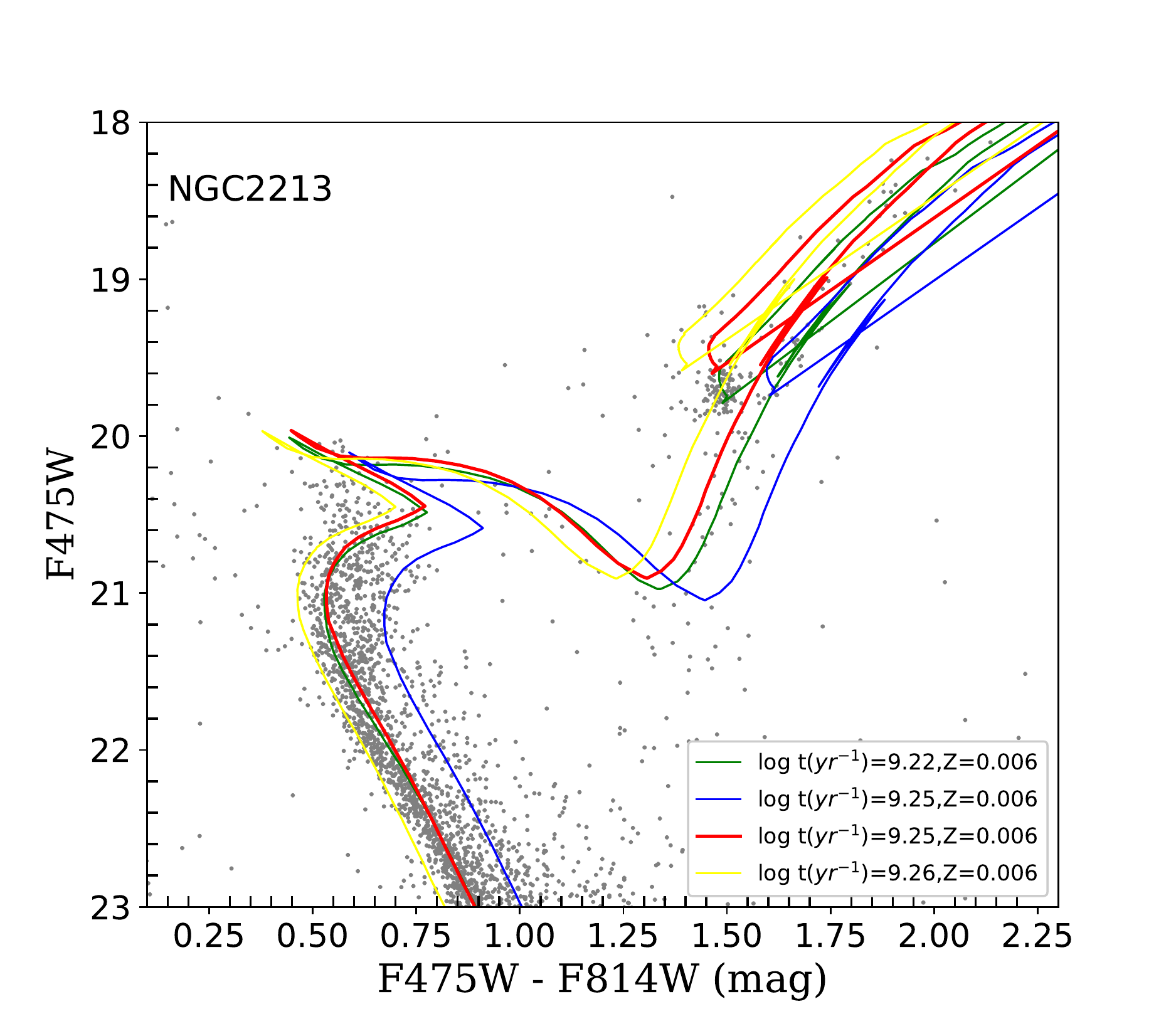}
\caption{CMDs of NGC 411 (top row), NGC 1718 (bottom left) and NGC 2213 (bottom right) along with the best-fitting isochrones, labelled with their ages.}
\label{pics:blablabla}
\end{figure*}

In this paper, we explore the question of whether clusters as young as $\sim$1.6 Gyr show evidence of MPs among their RGB stellar populations. We organise the paper as follows. In Section 2, we discuss the observational data as well as our data reduction and analysis procedures. In Section 3, we present our results and discuss their astrophysical implications. Section 4 summarises our results and conclusions.

\section{Observations and Methods}

\subsection{Data and Photometry}

We used the {\sl Hubble Space Telescope}'s ({\sl HST}) Advanced Camera for Surveys (ACS)/Wide Field Camera (WFC) and  {\sl HST}/Wide Field Camera 3 (WFC3) Ultraviolet--Visible channel (UVIS) archival data obtained from the {\sl HST} Legacy Archive. We downloaded WFC3/UVIS images obtained through the F475W and F814W filters for all three of our sample clusters (Proposal ID: 12257; Principal Investigator: L. Girardi).

The NGC 411 data set also contained a set of images taken through the F336W WFC3/UVIS filter, in addition to the F475W and F814W observations. Exposure times for the long-exposure images were 2200 s, 1520 s and 1980 s in the F336W, F475W and F814W filters, respectively. The long-exposure images of NGC 1718 had exposure times of 1440 s for F475W and 1430 s for F814W, whereas the long-exposure images of NGC 2213 had exposure times in F475W and F814W of 1440 s and 1430 s, respectively.

The data set used in this paper is summarised in Table 1.

To perform point-spread-function (PSF) photometry on the observations of the three clusters, we used two independent photometric software packages, IRAF/DAOPHOT \citep{1987PASP...99..191S} and DOLPHOT (Dolphin 2000). Initial analyses using both packages produced mutually consistent results. We henceforth show our results, in the VegaMag photometric system, based on application of DOLPHOT. We referred to the DOLPHOT manual\footnote{http://americano.dolphinsim.com/dolphot/dolphotWFC3.pdf} for the relevant steps to perform our photometric analysis. Briefly, the following are the recommended steps for WFC3 filters: 

\begin{figure*}
\centering 
\includegraphics[width=1.0\textwidth]{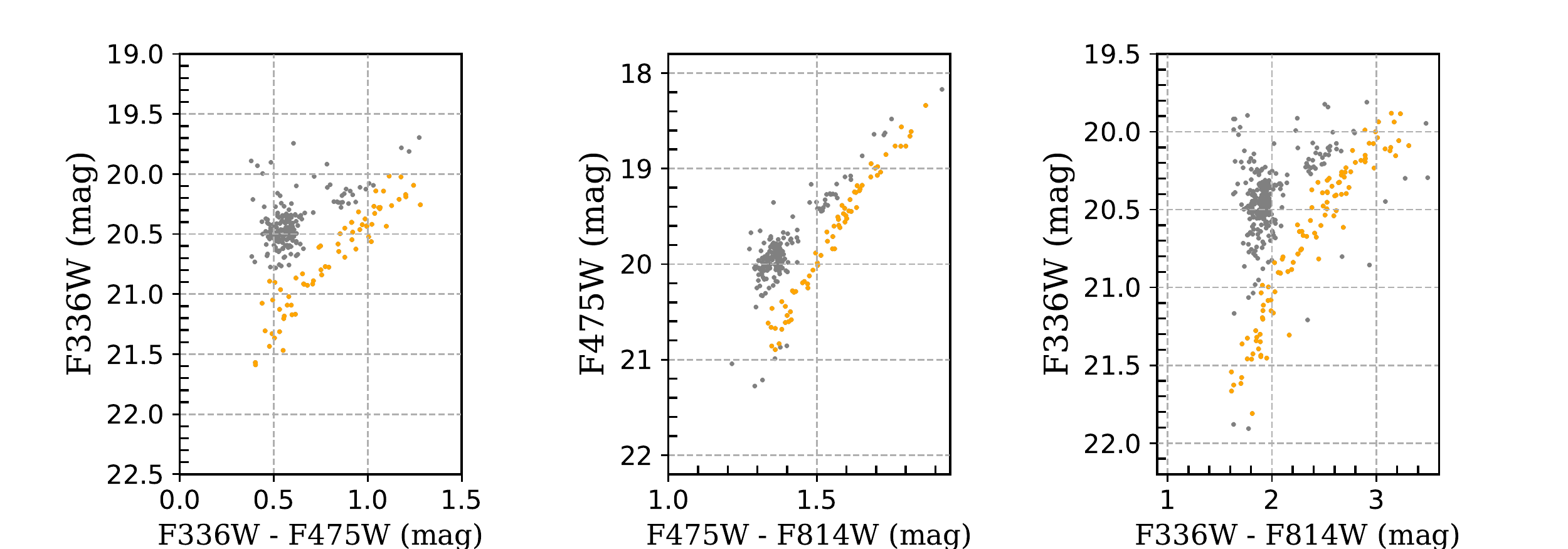}\\
\medskip
\includegraphics[width=1.0\textwidth]{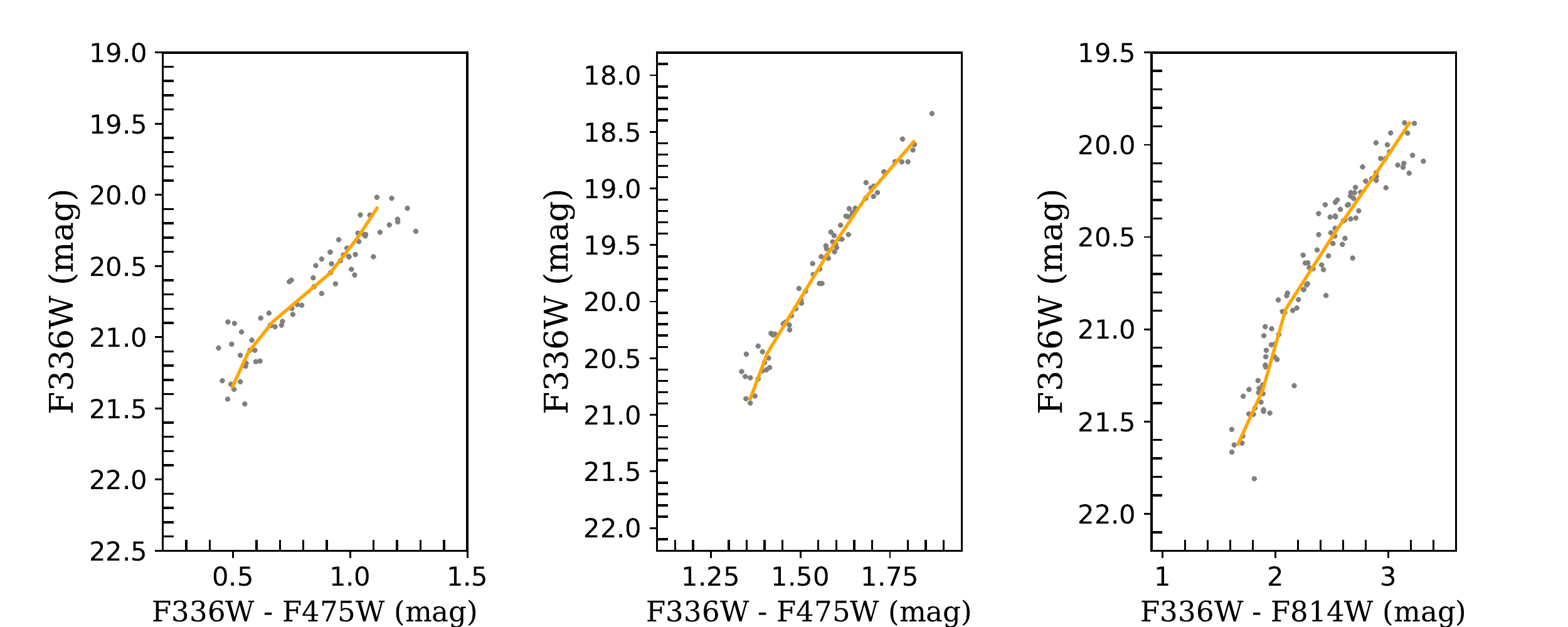}
\medskip
\includegraphics[width=.3\textwidth]{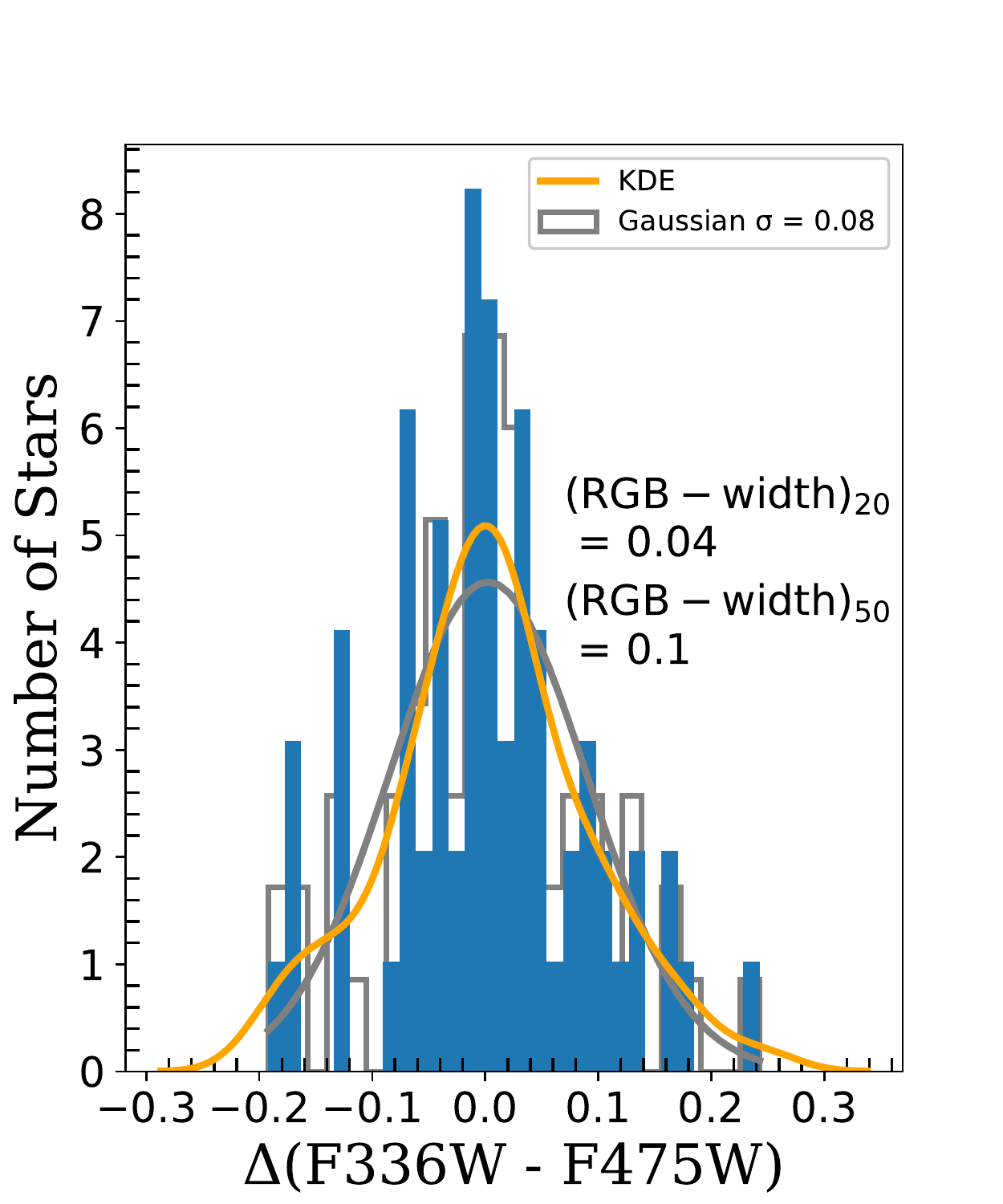}\quad
\includegraphics[width=.3\textwidth]{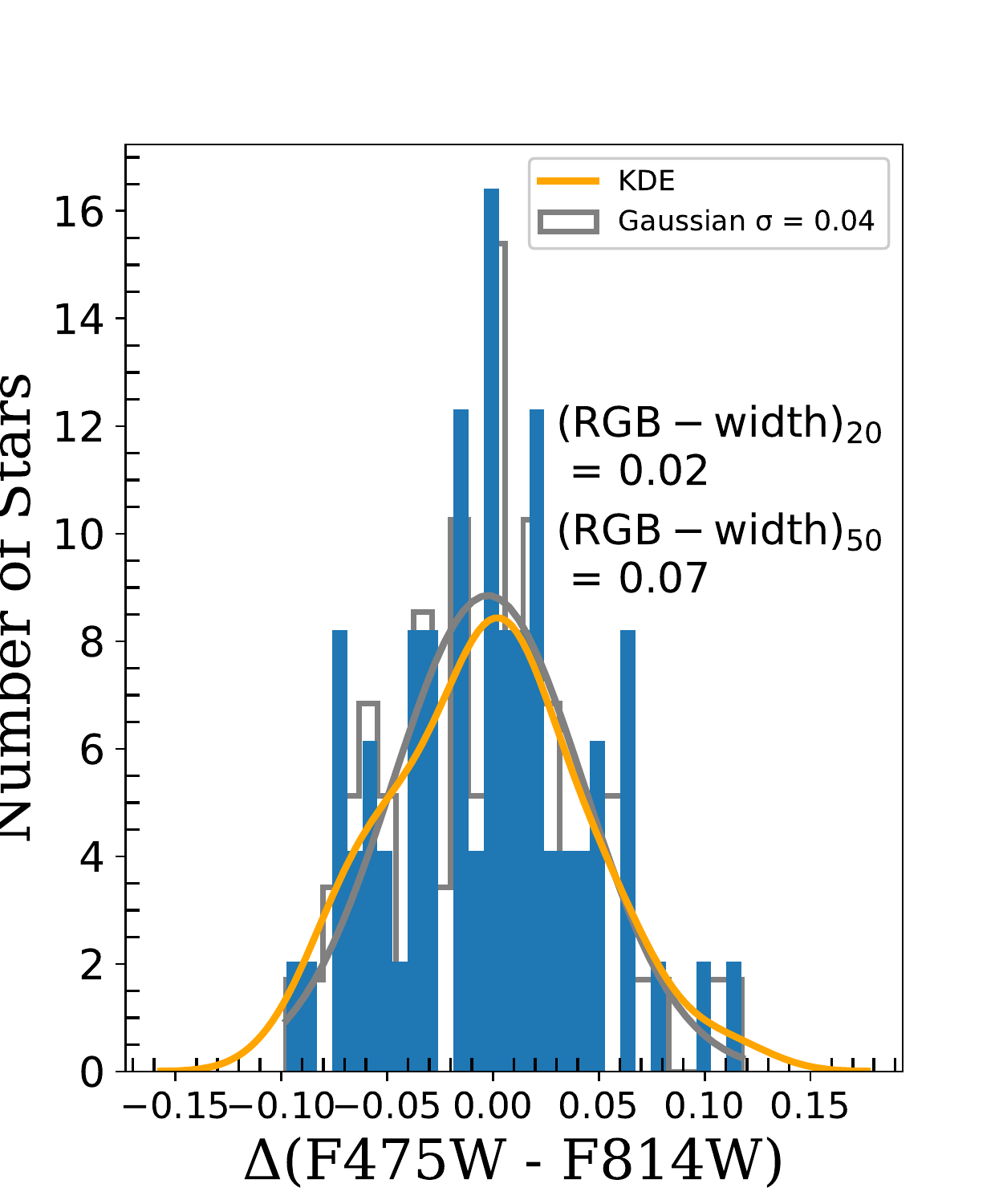}\quad
\includegraphics[width=.3\textwidth]{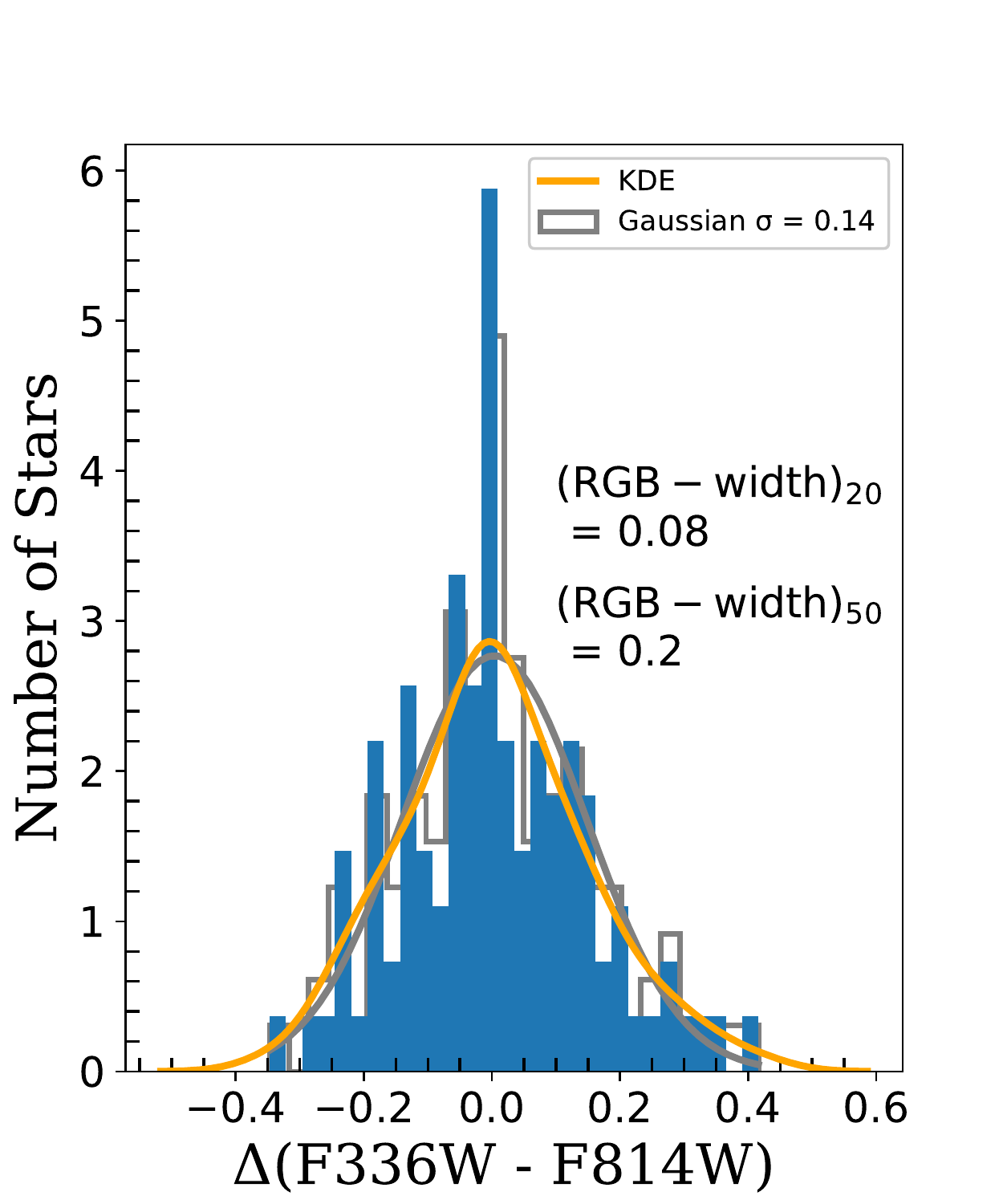} \\
\caption{RGB analysis of NGC 411. (top row) Selected RGB (orange) and red clump/AGB members (grey). (middle row) Best-fitting fiducial lines drawn through the RGB population after removal of the red clump stars. (bottom row) Gaussian probability distribution functions (PDFs) and the corresponding kernel density estimations (KDEs). The RGB widths at 20 per cent and 50 per cent of the maxima of the Gaussian distributions are superimposed.}
\label{pics:blablabla}
\end{figure*}

\begin{itemize}
    \item Mask bad pixels using the command {\tt WFC3MASK};
    \item Use the {\tt SPLITGROUPS} command to split the images according to the chips used for the observations;
    \item Apply {\tt CALCSKY} to calculate the sky brightness level;
    \item Perform PSF photometry using the command {\tt DOLPHOT}. We applied our photometric approach to all long-exposure \texttt{.flt} images simultaneously, using the drizzled (\texttt{.drz}) filter image as our reference image, for each star cluster. We chose the drizzled image of the filter with the longest exposure time as our reference image;
    \item Combine all catalogues obtained from the different chips, for each filter.
\end{itemize}

We selected cluster stars from the output catalogue based on their sharpness parameter, defining an operational range of [$-$0.4, 0.4]. The total number of objects thus obtained consisted of 113,111 stars for NGC 411, 94,038 stars for NGC 1718 and 93,227 stars for NGC 2213. We next matched the catalogues obtained from images taken with different exposure times. For NGC 411, we selected stars with magnitudes $m_{\rm F336W} < 18.2$ mag, $m_{\rm F475W} < 18.0$ mag and $m_{\rm F814W} < 18.1$ mag. In NGC 1718, we selected stars with $m_{\rm F475W} < 18.2$ mag and $m_{\rm F814W} < 18.1$ mag, whereas for NGC 2213 we selected stars with $m_{\rm F475W} < 18.5$ mag or $m_{\rm F814W} < 18.9$ mag. For our final catalogue of NGC 411, we combined all stars from the long-exposure photometry. We cross-identified the stars found in different filters in all three clusters using the same procedure. The matched stellar catalogues with magnitudes in all three (NGC 411) photometric bands contain in total, respectively, 117, 111 and 104 RGB stars in three colour combinations ($m_{\rm F336W} - m_{\rm F475W}$, $m_{\rm F475W} - m_{\rm F814W}$ and $m_{\rm F336W} - m_{\rm F814W}$, respectively). For NGC 1718 and NGC 2213, we selected the stars based on visual inspection. 

\begin{table}
	\centering
	\caption{Data set analysed in this paper}
	\label{Table 1}
	\begin{tabular}{lccccr} 
		\hline
		Cluster &Camera &Exposure & Filters \\
		        &       &time (s) &         \\
        
		\hline
		NGC  411 & WFC3/UVIS & 2200 & F336W &  \\
		         &           & 1520 & F475W &  \\
		         &           & 1980 & F814W &  \\
		        
		NGC 1718 & WFC3/UVIS & 1440 & F475W &  \\
		         &           & 1430 & F814W &  \\
		         
		NGC 2213 & WFC3/UVIS & 1440 & F475W &  \\
		         &           & 1430 & F814W &  \\
		\hline
	\end{tabular}
\end{table}

\begin{figure*}
\centering
\includegraphics[width=.4\textwidth]{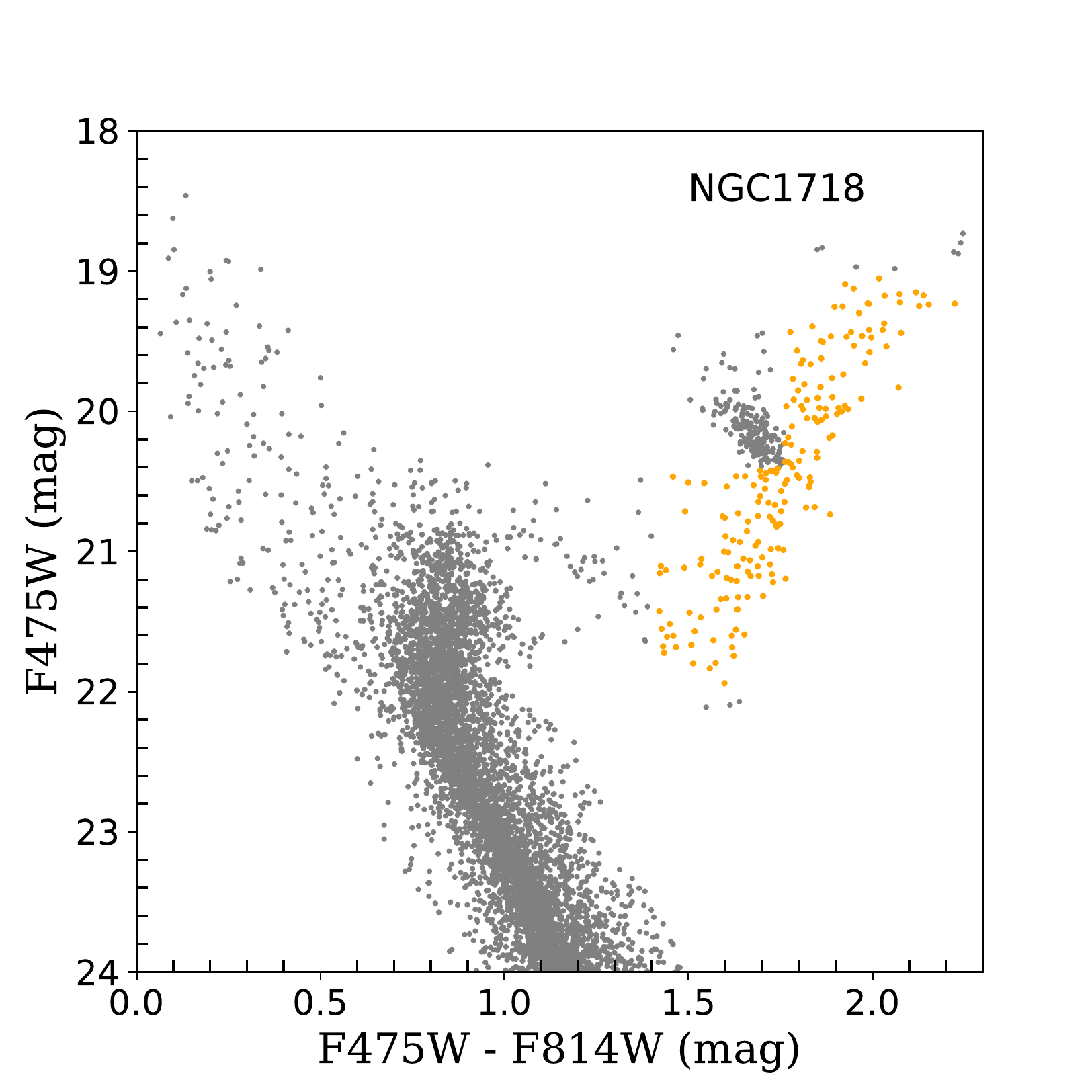}\quad
\includegraphics[width=.4\textwidth]{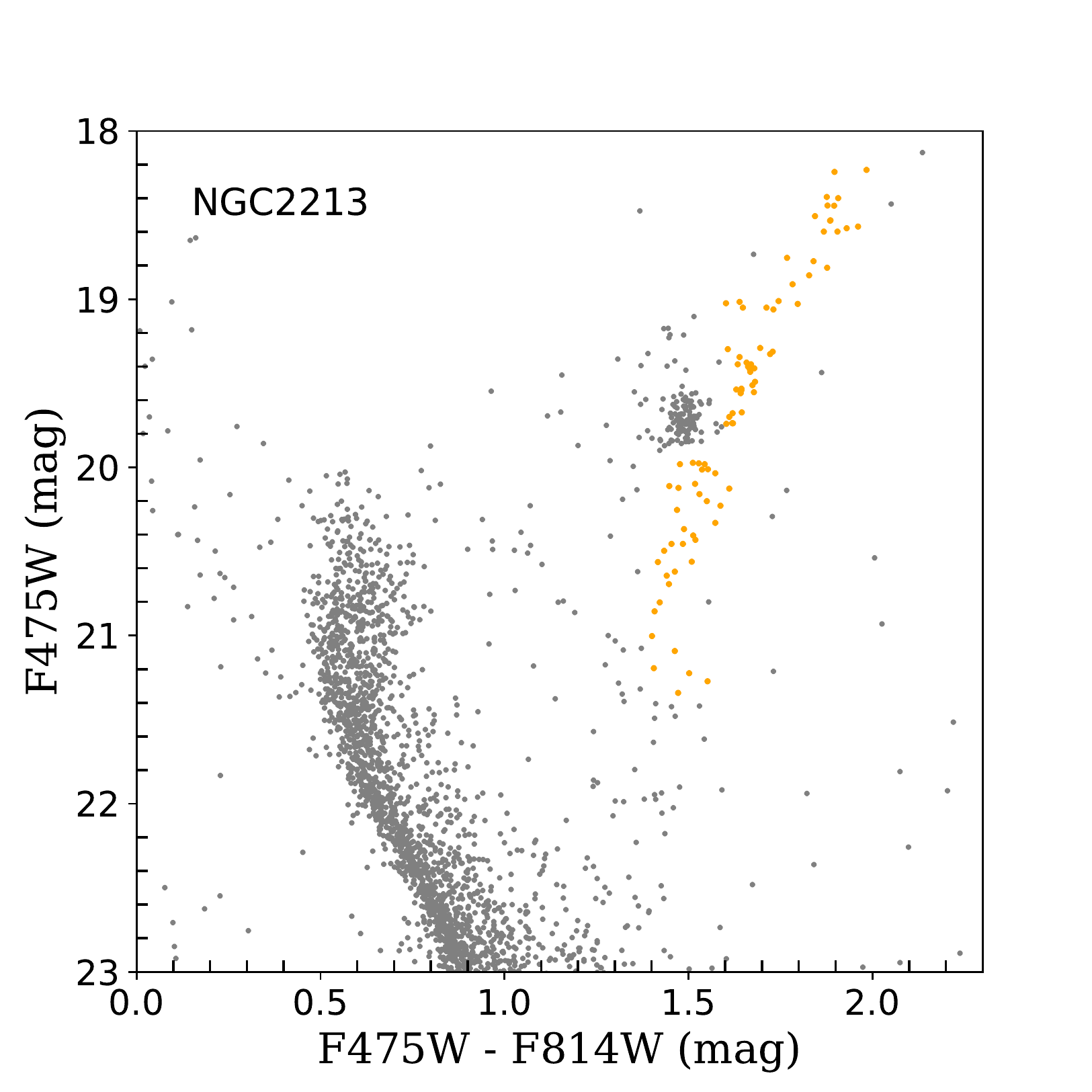}
\medskip
\includegraphics[width=.4\textwidth]{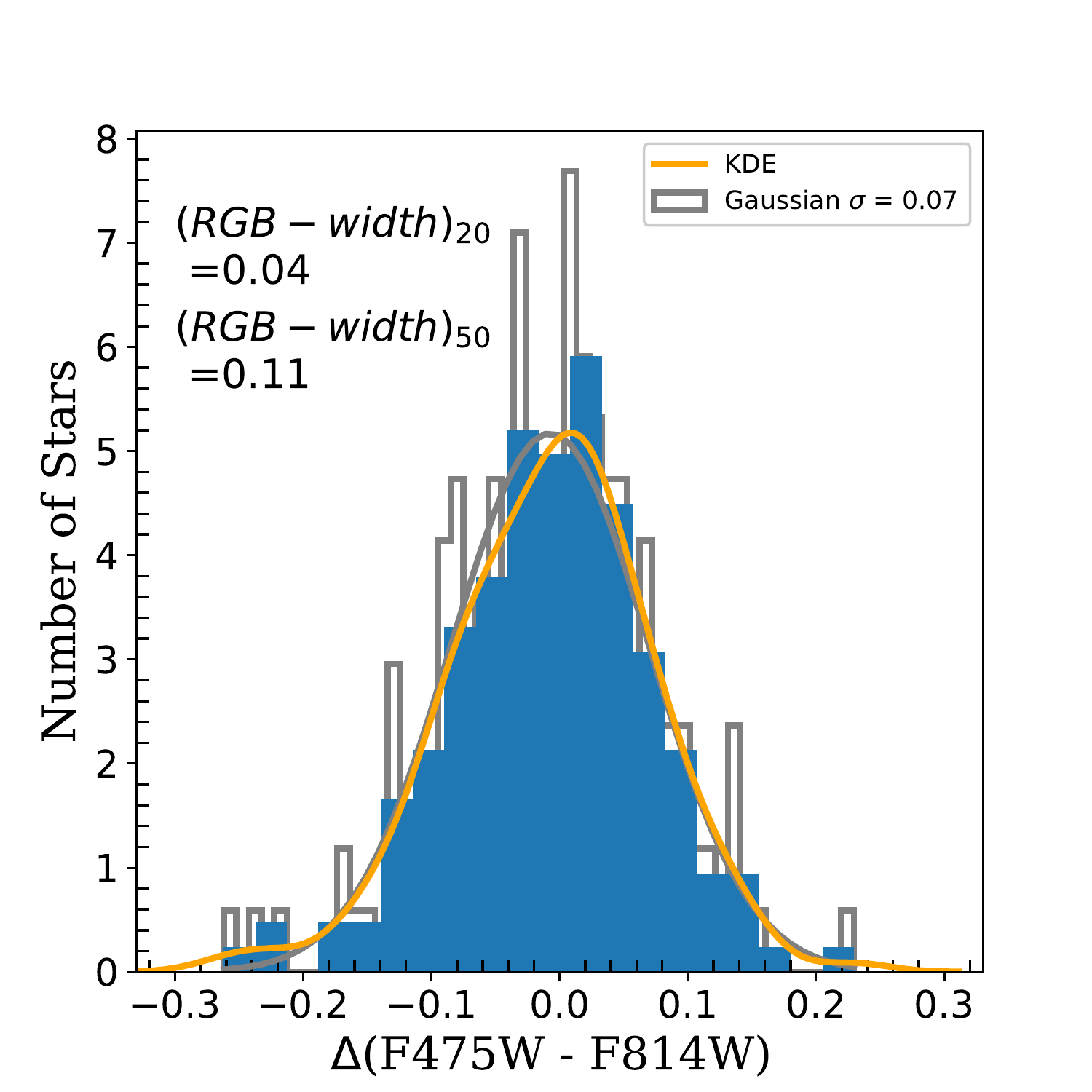}\quad
\includegraphics[width=.4\textwidth]{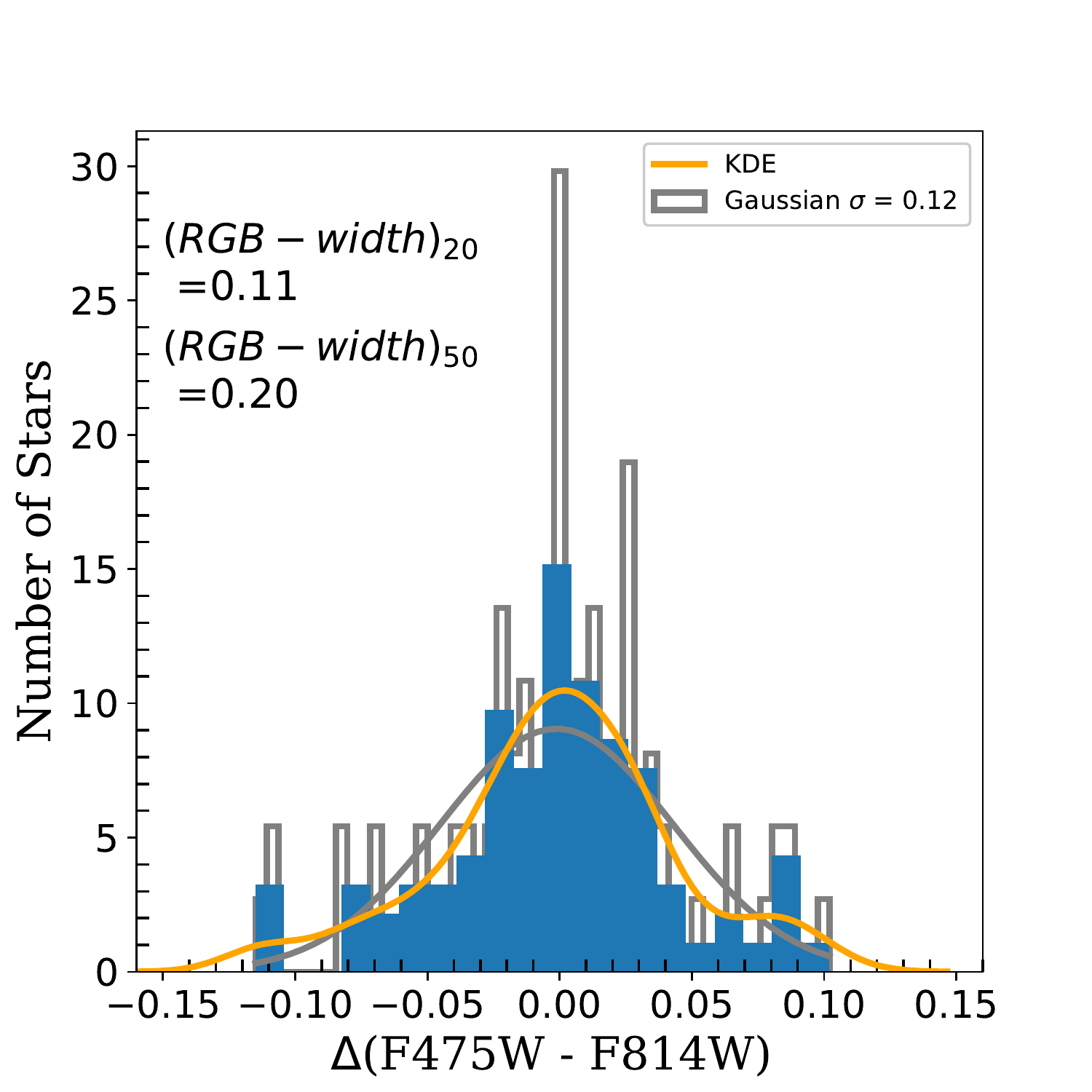}
\caption{Selected RGB samples for NGC 1718 and NGC 2213, along with their CMDs. The Gaussian PDFs are shown, and the means and standard deviations are given. The respective KDEs have been superimposed.}
\label{pics:blablabla}
\end{figure*}

\subsection{Corrections for differential reddening}

To remove any effects caused by potential spatially variable foreground reddening, differential reddening corrections were applied. \citet[][their fig. 8]{681408180} provide an example to highlight the importance of removing such reddening effects. Their results show that clear differences are seen in the widths of both the MSs and RGBs of several clusters upon application of these corrections.

We therefore followed the procedure of \citet[][their section 3.1]{681408180} to correct our photometry and remove any effects of differential reddening: 

\begin{itemize}
    \item We rotated the CMD such that the abscissa is parallel to the direction of the expected differential reddening. The rotation angle was calculated using the ratio of the extinction effects in the two filters which were used to generate the CMD. 
    
    \item Since we focus on the clusters' RGB stars, we selected our reference stars from the RGB sample set. By fitting a central fiducial curve to this region in CMD space, we proceeded to calculate the distance from the fiducial curve to each star. 
    
    \item  The associated reddening value was determined by averaging the reddening values of their $N$-nearest reference stars. We adopted $N=50$ \citep{681408180}.
\end{itemize} 

\begin{figure*}
\centering
\includegraphics[width=.4\textwidth]{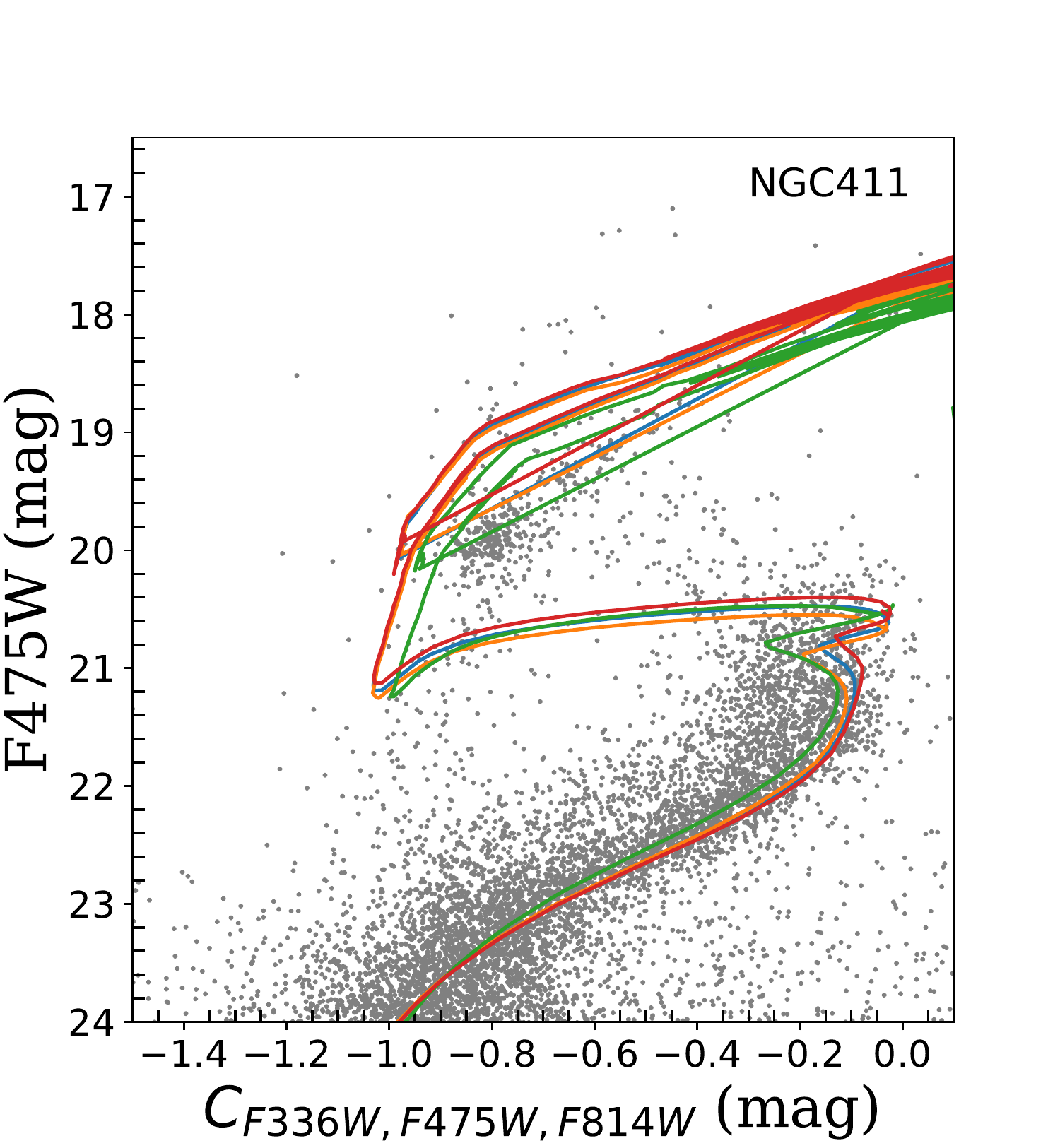} 
\includegraphics[width=.4\textwidth]{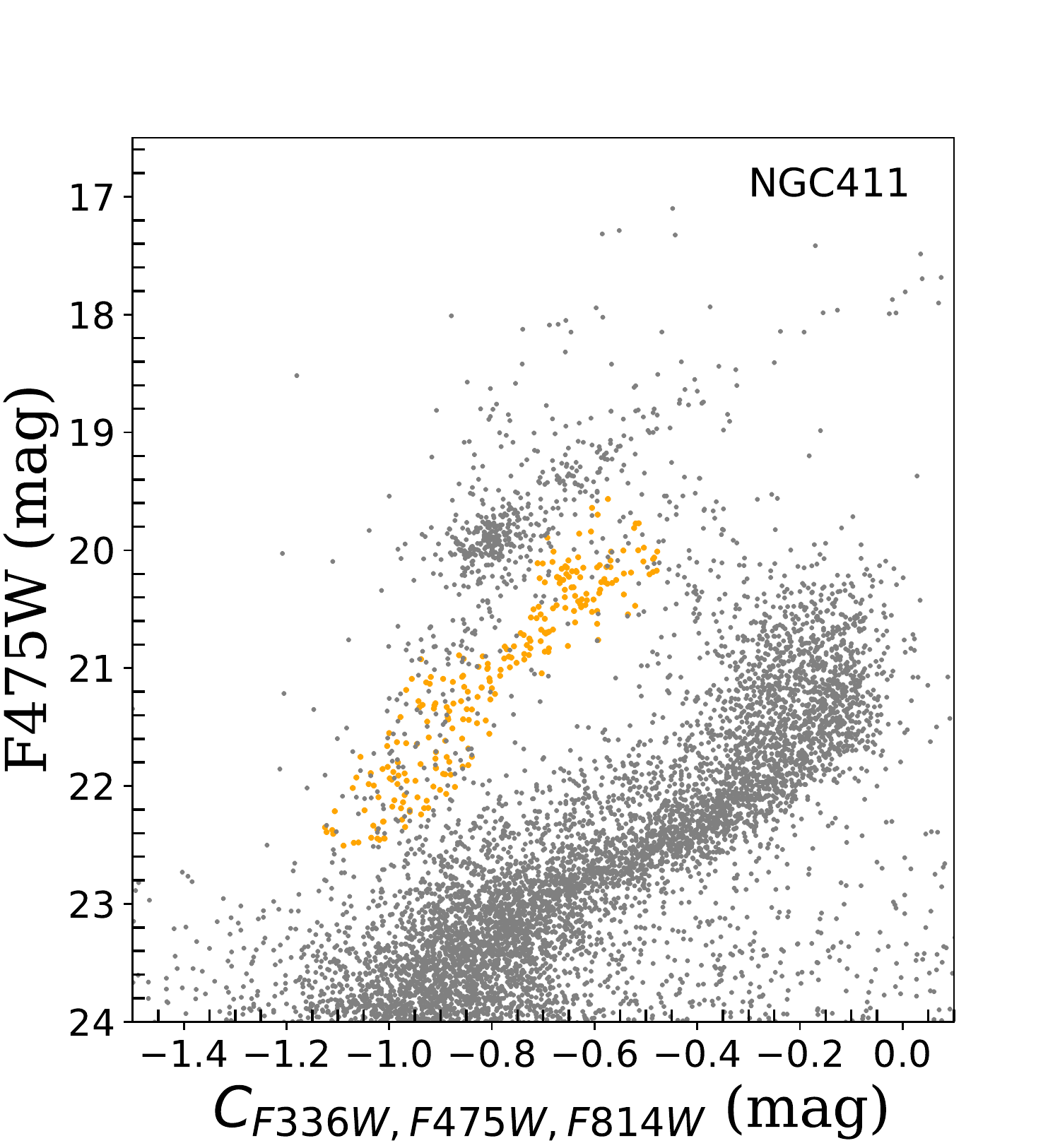} \\
\medskip
\includegraphics[width=.4\textwidth]{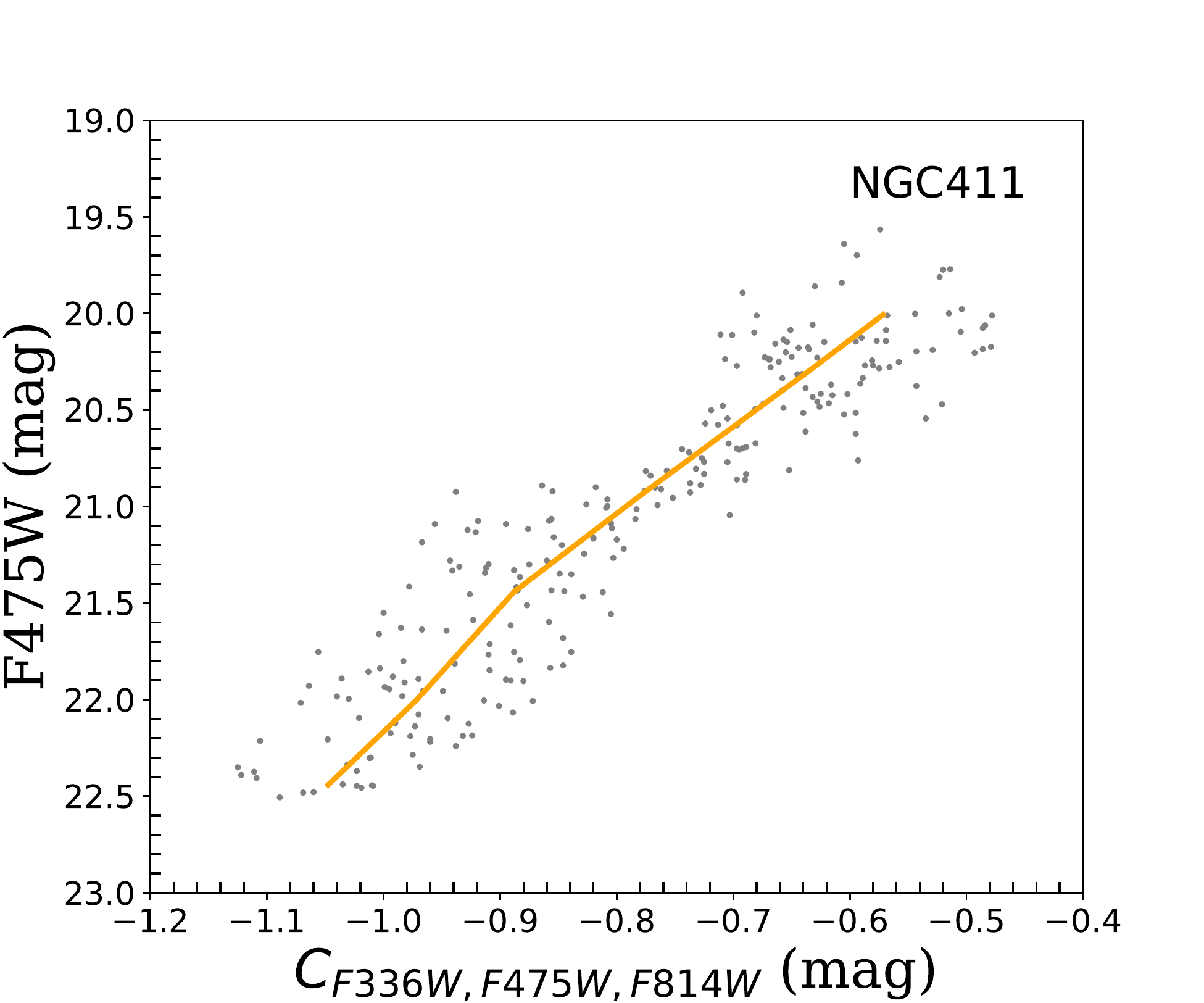}
\includegraphics[width=.4\textwidth]{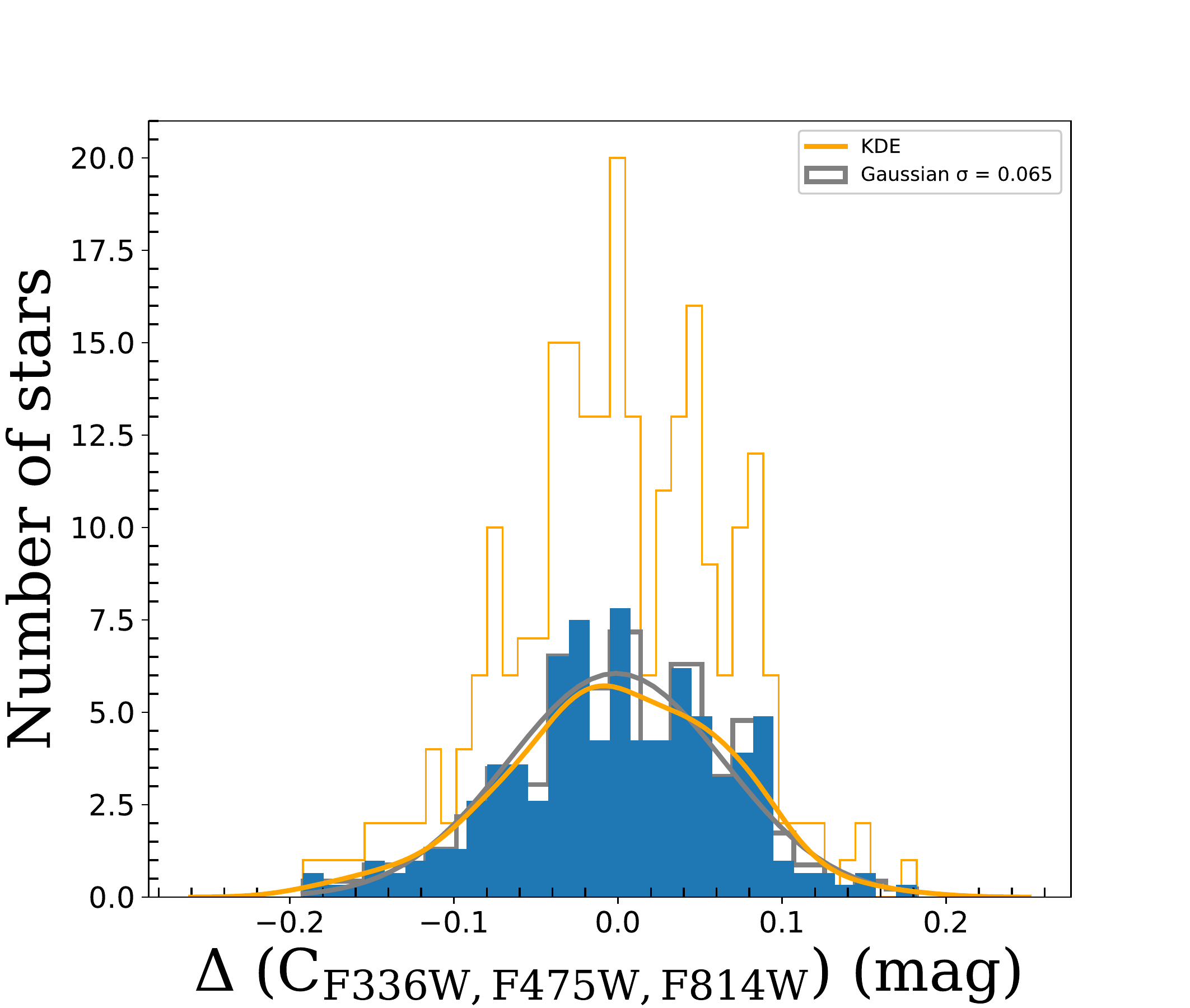} \\
\caption{$C_{\rm F336W, F475W, F814W}$ versus F475W CMD of NGC 411, with isochrones overplotted (top left). Parameters used for the isochrones: $\log( t \mbox{ yr}^{-1}) = 9.18, 9.20, 9.22; Z = 0.002, 0.003; A_V = 0.25$ mag. In the top right-hand panel, orange dots show the RGB stars selected for this analysis. The bottom left-hand panel displays the fiducial curve drawn through the RGB sample. In the bottom left-hand panel, the Gaussian PDF is shown as the solid grey curve (with the distribution's mean and standard deviation $\sigma$ calculated from the RGB data), while the orange curve represents the KDE.}
\label{pics:blablabla}
\end{figure*}
The average change in colour, $\delta$, owing to differential reddening in $m_{\rm F336W - \rm F475W}$ was $<$ 0.01 mag, $\delta (m_{\rm F475W - \rm F814W})$ $<$ 0.02 mag and $\delta (m_{\rm F336W - \rm F814W})$ $<$ 0.01 mag for NGC 411. For both NGC 1718 and NGC 2213, the change in colour was $\delta (m_{\rm F475W} - m_{\rm F814W}) < 0.02$ mag. Since the measured differential reddening effects are negligible and do not affect our results, we did not implement reddening corrections. 

\subsection{RGB stars}

We selected stars with luminosities equal to or brighter and redder than the red clump (RC). If we include bluer stars, there is a real possibility that our sample set might be contaminated by AGB stars. For NGC 411, we adopted the magnitude ranges $20.1 < m_{\rm F336W} < 21.9$ mag, $19.9 < m_{\rm F475W} < 22.0$ mag and $20.1 < m_{\rm F814W} < 21.75$ mag. For NGC 1718, the selected stellar magnitudes covered the range $18.1 < m_{\rm F475W} < 24.2$ mag and $18.03 < m_{\rm F814W} < 25.23$ mag, whereas for NGC 2213 we adopted $18.5 < m_{\rm F475W} < 20.8$ mag and $18.3 < m_{\rm F814W} < 21.2$ mag. These boundaries also define the magnitude ranges from the bottom to the tip of the RGB, not including RC stars. 
   
For NGC 411, the colour ranges we adopted for our RGB sample stars were $\Delta (m_{\rm F336W} - m_{\rm F475W}) = 0.80$ mag, $\Delta (m_{\rm F45W} - m_{\rm F814W}) = 0.50$ mag and $\Delta (m_{\rm F336W} - m_{\rm F814W}) = 0.98$ mag. For NGC 1718 and NGC 2213, we adopted  $\Delta(m_{\rm F475W} - m_{\rm F814W}) = 0.50$ mag. We selected only those RGB stars which were well separated from the respective RC of each cluster. 

\subsection{Field-star decontamination}

To calculate the width of the RGB, we must ensure that our RGB sample is free from any contamination by field stars. Our decontamination process is as follows.   
\begin{itemize}
    \item We selected a field-dominated circular region well away from the cluster's boundary region. The centre of NGC 411 is located at R.A. (J2000) = $01^{\rm h}$$07^{\rm m}$$55.673^{\rm s}$, Dec. (J2000) = $-71^{\circ}46'06.06''$ and we adopted a radius of $60''$, or $3 \times$ the cluster's core radius \citep{2016Natur.529..502L}, for the decontamination process;
    \item We applied the same RGB selection criteria to the cluster and field CMDs so as to extract a `field RGB' sample;  
    \item We statistically subtracted the field component from the cluster's RGB sample. 
\end{itemize}
We followed the same procedure for our other two clusters. The cluster centres of NGC 1718 and NGC 2213 are located at R.A. (J2000) = $04^{\rm h}$$52^{\rm m}$$25.387^{\rm s}$, Dec. (J2000) = $-67^{\circ}03'02.38''$ \citep{Goudfrooij_2014} and R.A. (J2000) = $06^{\rm h}$$10^{\rm m}$$41.94^{\rm s}$, Dec. (J2000) = $-71^{\circ}10'42.38''$ \citep{Baumgardt_2013}, respectively. We considered radii of $60''$ and $75''$, equivalent to twice and 2.5 times the clusters' core radii, respectively \citep{Goudfrooij_2014}. 
\begin{table*}
	\centering
	\caption{Isochrone parameters used to match the three clusters studied in this paper}
	\label{Table 2}
	\begin{tabular}{lcccccc} 
		\hline
		Cluster &$\log t$    & $M_{\rm cluster}$  & $Z$ & $A_V$  & $(m - M)_0$  & Reference\\
		        &[yr$^{-1}$] & $(10^4$ M$_\odot$) &     & (mag)  &  (mag) \\
        
		\hline
		NGC 411 & 9.22 & 4.24 & 0.003 & 0.25 & 19.05 & 1  \\
		        & 9.20 & 4.22 & 0.002 & 0.25 & 18.56 & 2  \\
		        & 9.18 & 4.23 & 0.003 & 0.25 & 19.05 & 3  \\
		        & 9.20 & 4.22 & 0.002 & 0.25 & 19.05 & 3  \\
\\
		NGC 1718 & 9.23 & 4.67 & 0.008 & 0.25 & 18.73 & 4 \\
		         & 9.27 & 4.64 & 0.008 & 0.25 & 18.73 & 5 \\
		         & 9.25 & 4.64 & 0.008 & 0.25 & 18.73 & 3 \\
		         & 9.25 & 4.64 & 0.01  & 0.25 & 18.73 & 3 \\
\\		        
		NGC 2213 & 9.22 & 4.83 & 0.006 & 0.25 & 18.45 & 2 \\
		         & 9.25 & 4.83 & 0.006 & 0.25 & 18.50 & 6 \\
		         & 9.25 & 4.83 & 0.006 & 0.19 & 18.36 & 3 \\
		         & 9.26 & 4.83 & 0.006 & 0.19 & 18.36 & 3 \\ 
		\hline
	\end{tabular}
	\flushleft
	References: (1) \cite{10.1093/mnras/stw1491}; (2) \cite{Goudfrooij_2014}; (3) This paper; (4) \cite{2017MNRAS.467.1112S}; (5) \cite{2016MNRAS.463.1632P}; (6) \cite{10.1093/mnras/sty580}.
\end{table*}
\section{Analysis}

\subsection{Isochrone fitting}

For NGC 411, we adopted an average extinction of $A_V = 0.25 \pm 0.01$ mag and a metallicity $Z = 0.006 \pm 0.003$ \citep{Li2016StellarClusters}. We used these values to explore the performance of a set of PARSEC stellar isochrones \citep{Goudfrooij_2014}, adjusting their metallicity $Z$ and age $t$ to visually search for the best-matching isochrone. We obtained as best-fitting parameters, $\log(t \mbox{ yr}^{-1}) = 9.22$ and $Z = 0.006$. The resulting isochrone is shown in Fig. 2. Based on a careful comparison with results from other studies, we are confident that our best-fitting metallicity, $Z = 0.006$, is most appropriate; we carefully examined the match of our best-fitting isochrone to the cluster’s RC. For the best-fitting age of $\log(t \mbox{ yr}^{-1}) = 9.22$, the resulting metallicity, $Z = 0.006$, is clearly the best fit. All best-fitting isochrone parameters for our three sample clusters are listed in Table 2.   

For NGC 1718 and NGC 2213, we compared PARSEC isochrones \citep{675688505} to our data so as to obtain an estimate of the clusters' ages, metallicities and distance moduli. We also compared our best-fitting age for NGC 1718 with the parameters derived by \cite{Goudfrooij_2014}, whereas for NGC 2213 we compared our parameters with those from \cite{10.1093/mnras/sty580}. Our best-fitting parameters for NGC 1718 are $\log( t \mbox{ yr}^{-1}) = 9.25 \pm 0.02$, $Z = 0.01 \pm 0.03$ and $(m - M)_0 = 18.73$ mag. For NGC 2213, we determined $\log(t \mbox{ yr}^{-1}) = 9.26$ ($\sim 1.81$ Gyr), $Z = 0.006$, $A_{V} = 0.19$ mag and $(m - M)_0 = 18.36$ mag. Comparing this result with other studies (see Section 4), we believe that our best-fitting age is indeed accurate. This best-fitting isochrone has been overplotted on the CMD of NGC 2213 in Fig. 2.

We have also estimated the level of possible helium-abundance variations in the three clusters. For  NGC 411, we derived a helium-abundance variation from the BaSTI theoretical stellar models \footnote{http://basti.oa-teramo.inaf.it}, following \citep{2018MNRAS.481.5098M}, of $\delta Y=0.003 \pm 0.001 (Y=0.300)$. For NGC 1718 and NGC 2213, $\delta Y = 0.002 \pm 0.001 (Y=0.350)$ and $0.004 \pm 0.002 (Y=0.300)$, respectively. The RGBs in our cluster sample are narrow, and so the maximum possible helium-abundance variation allowed is also very small. Using the same approach, we  determined an upper limit to the nitrogen-abundance variation in NGC 411 of $\Delta$[N/Fe] = 0.3 dex. Here, we used the F336W filter, combined with F475W and F814W, since it is most sensitive to nitrogen variations. For NGC 1718 and NGC 2213, the available data do not allow us to determine useful upper limits.

\subsection{RGB analysis}

For NGC 411, we plotted the RGB sample in the $(m_{\rm F336W} - m_{\rm F475W})$ versus $m_{\rm F336W}$, $(m_{\rm F475W} - m_{\rm F814W})$ versus $m_{\rm F475W}$, and $(m_{\rm F336W} - m_{\rm F814W})$ versus $m_{\rm F336W}$ CMDs: see Fig. 3 (first panel). We removed the RC stars from our RGB sample in the second panel of Fig. 3. 
We then calculated the mean and standard deviation ($\sigma$) of these colours and obtained the Gaussian probability density function (PDF), as shown by the grey solid curves in the bottom panel of Fig. 3. We derived kernel density estimates (KDEs) from the Gaussian PDFs for the three colour combinations. The resulting KDEs have been superimposed on the histograms of the RGB data in the final panel of Fig. 3. There are no significant differences between the Gaussian PDFs and the KDE distributions of these colours. Plotting KDE distributions on Gaussian PDFs can be a helpful tool to detect if there are any hidden second peaks smoothed by the Gaussian PDF. Neither our Gaussian PDFs nor our KDE distributions reveal any differences in either distributions. We followed the same procedure for NGC 1718 and NGC 2213 (see Fig. 4).  

We also calculated the RGB widths at 20 per cent and 50 per cent of the distributions' maxima. For the (F336W, F475W) colour combination, the widths at 20 and 50 per cent of the distributions' maxima are RGB-width$_{20}$ = 0.04 mag and RGB-width$_{50}$ = 0.1 mag, respectively. For (F475W, F814W), RGB-width$_{20}$ and RGB-width$_{50}$ are 0.02 mag and 0.07 mag, respectively. The RGB-width$_{20}$ and RGB-width$_{50}$ values for (F336W, F814W) are 0.08 and 0.2, respectively; see Fig. 3 (bottom panel).

Next, for NGC 411, we selected RGB stars using three different CMDs and colours in different combinations. Using the ($m_{\rm F336W} - m_{\rm F474W}), (m_{\rm F475W} - m_{\rm F814W})$ and $(m_{\rm F336W} - m_{\rm F814W})$ colour combinations, we created the corresponding pseudo-colour indices $C_{\rm F336W, F475W, F814W}$ to check for any effects of possible MPs among the RGB populations in our clusters' CMDs. The pseudo-colour index, $C_{\rm index}$, is defined as $C_{\rm F336W, F475W, F814W} = (m_{\rm F336W} - m_{\rm F475W}) - (m_{\rm F475W} - m_{\rm F814W})$. We used this index to detect any split or spread in the RGB of NGC 411, by plotting the pseudo-CMD; however, we did not find any evidence of a split/spread in the RGB sections of the NGC 411 CMDs (see Fig. 5).

We used the same methodology to plot the Gaussian PDFs and KDEs pertaining to the pseudo-CMDs. Briefly, we selected the RGB sections in the pseudo-CMD and plotted them in the $m_{\rm F475W}$ versus $C_{\rm F336W, F475W, F814W}$ plane. Selected RGB stars are displayed in the top right-hand panel of Fig. 5. We proceeded to fit a Gaussian PDF to these data, superimposed with the corresponding KDE.
 
\subsection{Artificial-star tests}

To determine whether the measured full width of the RGB depends entirely on the photometric uncertainties or if there are any external parameters affecting the broadening of the RGB, we created artificial input stellar catalogues containing 180,000, 120,000 and 150,000 stars for use with, respectively, the NGC 411, NGC 1718 and NGC 2213 observational data. We used the {\tt -fakestar} option in DOLPHOT and reran the DOLPHOT photometric routines on the artificial stars, thus generating artificial CMDs for comparison with the observed cluster CMDs. We followed the following steps to generate the input artificial stellar catalogues: 
\begin{itemize}
    \item  We assigned random coordinates to the artificial stars based on a random selection of 104 lower-RGB stars from our observed sample of NGC 411 RGB stars, which were visually considered; 
    \item We added random perturbations to these coordinates to avoid spatial overlaps with any of the observed lower-RGB stars. These perturbations were equivalent to half the distance of any of the selected lower-RGB stars to their (spatially) closest lower-RGB star;
    \item We assigned magnitudes to the artificial stars based on the closest-matching segment of the best-fitting PARSEC isochrone. We interpolated this segment uniformly for the entire artificial-star catalogue;
    \item Finally, we matched the coordinates randomly with the resulting magnitudes to complete our input artificial RGB star catalogue.
\end{itemize}

For NGC 411, we recovered 162,443 of the 180,000 stars entered as input stars in DOLPHOT. We recovered 102,324 and 127,453 of, respectively, the 120,000 and 150,000 input artificial stars for the other two clusters. We subsequently analysed  the magnitudes of the observed and artificial RGB sequences versus their colour index, $C_{\rm index}$. 

We use $C_{\rm index}$ as an important diagnostic to identify the possible presence of MPs in our young and intermediate-age star clusters. Second-generation stars are expected to have higher He and N abundances, as well as lower carbon (C) abundances, than first-generation stars. Their $(m_{\rm F336W} - m_{\rm F814W})$ colours will hence become bluer owing to the increased temperatures supported by the higher He abundance. In contrast, the $(m_{\rm F336W} - m_{\rm F475W})$ colours will become redder owing to an N absorption band found within the F336W filter bandpass, which is enhanced by the higher N abundance. Therefore, these differences between both generations of stars in a single star cluster are accentuated in the $C_{\rm F336W, F475W, F814W}$ combination, which is thus a good choice to probe for possible chemical abundance variations associated with MPs \citep{675688540}.

Since we added the artificial stars to our science images directly and used the same recovery and analysis techniques as for our science sample, our artificial stellar sample is subject to the same crowding effects and sampling incompleteness as our science catalogue. Hence, we can directly compare the shapes -- particularly the widths -- of the observed and artificial RGBs. The recovery rates for the artificial star tests were high ($\sim 85$--90\%), and the $C_{\rm index}$ distributions of both observed and artificial data were essentially the same, indicating that our results are unlikely significantly impacted by incompleteness or other observational effects. 

\section{Discussion and Conclusions}

In this paper, we have performed a photometric analysis of archival {\sl HST} images of the young, $<2$ Gyr-old SMC cluster NGC 411 and the similarly aged LMC clusters NGC 1718 and NGC 2213. We compared the CMDs of our cluster sample with both theoretical isochrones and parameter determinations from the literature (see below). We generated PARSEC isochrones to compare our observational results with the theoretical models and literature values. Our best-fitting values are slightly different from those previously published in the literature; they are listed in Table 2. 

We base our results on three colour combinations for NGC 411 ($m_{\rm F336W} - m_{\rm F475W}, m_{\rm F475W} - m_{\rm F814W}$ and $m_{\rm F336W} - m_{\rm F814W}$) and one colour ($m_{\rm F475W} - m_{\rm F814W}$) for both NGC 1718 and NGC 2213. For NGC 411, we used the pseudo-colour index $C_{\rm F336W, F475W, F814W}$ and pseudo-CMDs to check for the presence of any MPs among the clusters' RGB population. Chemical dispersions are expected to manifest themselves as a difference in the spread in the $m_{\rm F475W}$ versus $C_{\rm F336W, F475W, F814W}$ pseudo-CMD. If MPs were present in our sample clusters, we would have expected a significant spread in the three colours we used. However, we did not detect any evidence of chemical variations or abundance spreads in NGC 411. The $C_{\rm F336W, F475W, F814W}$ pseudo-colour index is sensitive to stellar He and N abundances, which are two of the most representative elements that tend to exhibit large abundance variations in star clusters \citep{2013MNRAS.431.2126M}. 

The SGB of NGC 411 has been investigated by \cite{10.1093/mnras/stw1491}. The latter authors show that the SGB morphology is consistent with the SSP scenario. A detailed analysis of the RGB of NGC 411 is performed for the first time in this paper. Analysis of the NGC 1718 CMD was first reported by \cite{refId1}. We compared our best-fitting age, metallicity and distance modulus with \cite{Goudfrooij_2014}. We conclude that our best-fitting isochrone parameters are slightly better for the RGB region. This is likely owing to the focus of \cite{Goudfrooij_2014} on NGC 1718's extended MS rather than its RGB. For NGC 2213, \cite{10.1093/mnras/sty580} provided estimates of the cluster's age, metallicity and distance modulus. Since our analysis focuses on the RGB stars, we obtained the best-fitting parameters pertaining to the clusters' RGBs. Our parameters of both age and distance modulus return a more robust fit compared with previously published results in the literature \citep{10.1093/mnras/sty580}, since those parameters were obtained to match the SGB of NGC 2213.

We calculated the widths of the RGBs using two complementary methods. One involved plotting the Gaussian PDFs and obtaining the standard deviations $\sigma$ for all three clusters. The second was based on calculating the RGB widths at 20 per cent and 50 per cent of the distributions' maxima. To indicate the possible presence of MPs, these widths should be broader than 2$\sigma$. The consistency of the observed RGB widths with that obtained from our artificial SSP indicates that our RGB samples are best represented by single-valued He and N abundances. We did not observe any RGB spreads in the three colour combinations. Therefore, we conclude that we did not detect any evidence of MPs in the RGB of NGC 411. We calculated the RGB spreads in three colours (see Fig. 3): $(m_{\rm F336W} - m_{\rm F475W}) = 0.08$ mag, $\sigma_{\rm (F475W - F814W)} = 0.04$ mag and $\sigma_{\rm (F336W - F814W)} =  0.14$ mag. 

We followed the same methodology for NGC 1718 and NGC 2213 using the $(m_{\rm F475W} - m_{\rm F814W})$ versus $m_{\rm F475W}$ CMD (see Fig. 4). The standard deviations are $\sigma_{\rm (F475W - F814W)} = 0.07$ mag and 0.12 mag for NGC 1718 and NGC 2213, respectively. The RGB widths at 20 per cent and 50 per cent are 0.04 mag and 0.11 mag (NGC 1718), and 0.11 mag and 0.24 mag (NGC 2213), respectively. On balance, we conclude once again that there is no clear evidence for the presence of chemical variations or MPs among the RGB populations in our cluster sample.

Observations of splits in the MS and at the eMSTO of young and (at least) moderately massive Magellanic Cloud star clusters have demonstrated the presence of complex stellar populations, suggestive of variations in their stellar rotation properties \citep{2018MNRAS.477.2640M}. The effects of stellar rotation on stellar populations have been observed in Galactic open clusters as well \citep{2018ApJ...863L..33M}. Nearly all of these young star clusters are sufficiently massive to host MPs among their MS stars. For example, NGC 1755, a young cluster with a mass of $\sim 10^4$ M$_\odot$ \citep{10.1093/mnras/stw608} or NGC 419, another young massive cluster with an age of 1.72 Gyr and a mass of $\sim 10^4$ M$_\odot$, are  comparably massive with respect to other star clusters which show clear evidence of MPs \citep{Martocchia2018TheNGC1978}. Thus, clusters with minimum masses of order $10^4$ M$_\odot$ tend to show evidence of MPs.
However, at present such a mass threshold has yet  to be explained theoretically. If we assume that mass is an important driver of MP formation, then NGC 411 should also have shown clear evidence of chemical abundance variations among its RGB stars. However, we have found no such evidence in NGC 411, which challenges the idea that mass may be the only or even the main driver of MP formation. The similarity of the observed RGB width in NGC 411 with that expected for an SSP suggests an absence of significant chemical abundance variations. In fact, for NGC 411, NGC 1718 and NGC 2213 we derive maximum possible helium-abundance variations of $\delta Y=0.003 \pm 0.001 (Y=0.300), 0.002 \pm 0.001 (Y=0.350)$ and $0.004 \pm 0.002 (Y=0.300)$, respectively. We  determined an upper limit to the nitrogen-abundance variation in NGC 411 of $\Delta$[N/Fe] = 0.3 dex, although the available data did not allow us to determine useful upper limits for our other sample clusters. Combined with similar results for NGC 419, NGC 1806 and NGC 1846 \citep{675688540}, our result is indeed inconsistent with mass being the {\em primary} driver of MP formation, although it seems likely that a sizeable minimum mass is still required. 

MPs appear ubiquitous in GCs older than 6 Gyr \citep{Milone2017Multiple1866}. For instance, NGC 339, an SMC cluster with an age of 6.3 Gyr and a mass of $8.3 \times 10^4$ M$_\odot$ hosts MPs \citep{Niederhofer2017The121}. However, we cannot directly compare MPs detected in young star clusters with those prevailing in old GCs; linking such young and old populations has thus far proven theoretically problematic. 

Moreover, our results for NGC 411, NGC 1718 and NGC 2213 are not consistent with age being the primary driver either \citep[cf. NGC 1978;][]{Martocchia2018TheNGC1978}. However, NGC 411 only represents a single data point, and so we should be careful not to overinterpret these results. Finally, at old ages the physical reality of genuine MPs in GCs may well be rather different from that at the younger ages of the clusters we have investigated in this paper. \cite{Milone2017Multiple1866} suggested that mass may be a driving factor of MP formation in old Milky Way GCs. They found a clear correlation of the RGB pseudo-colour width of the second-generation stars with their host clusters’ mass. The chemical abundance variations are well-established for old GCs; however at the younger ages (1--2 Gyr) discussed in this paper, we only have circumstantial evidence of the possible presence of MPs based on a broadening of distinct CMD features, and not of clearly distinct sequences.

Using {\sl HST} photometry, we have compared the pseudo-colour index dispersion of our observed RGB sequences with an artificial SSP, and we found that their widths are fully consistent with one another. We adopted a colour index which is sensitive to He and N abundance variations, so the observed consistency indicates the absence of significant He and N abundance variations. He and N are two elements whose abundances tend to vary most in star clusters in general. Our results thus imply that an SSP model is sufficient to explain the RGB broadening in our sample clusters.

Cluster age could be a determining factor driving the presence of MPs in LMC and SMC clusters. However, since our results do not reveal the presence of any MPs in some of the more massive Magellanic Cloud star clusters, the possibility of a combination of cluster age and mass affecting the presence of MPs cannot be ruled out. Nevertheless, to confirm which parameter may dominate the MP formation scenario, we will next apply our techniques to a larger sample of young and massive clusters.  

\section*{Data Availability}

The {\sl HST} data analysed in this paper are publicly available from the Hubble Legacy Archive (Proposal ID: 12257; Principal Investigator: L. Girardi). 

\section*{Acknowledgements}
This paper is based on observations made with the NASA/ESA {\sl HST} and obtained from the Hubble Legacy Archive, which is a collaboration of the Space Telescope Science Institute (STScI/NASA), the Space Telescope European Coordinating Facility (ST-ECF/ESA) and the Canadian Astronomy Data Centre (CADC/NRC/CSA). This work has made use of BaSTI web tools. We thank Chengyuan Li and Yujiao Yang for their practical assistance with aspects of the data reduction. S.K. acknowledges funding support from the International Macquarie Research Excellence Scheme (iMQRES).












\bsp	
\label{lastpage}
\end{document}